\documentclass[]{jfm}

\usepackage{graphicx}
\usepackage{newtxtext}
\usepackage{newtxmath}
\usepackage{natbib}
\usepackage{hyperref}
\usepackage{subfigure}
\usepackage{booktabs}
\usepackage{color}
\usepackage{multirow}
\hypersetup{
    colorlinks = true,
    urlcolor   = blue,
    citecolor  = black,
}

\newcommand{\RomanNumeralCaps}[1]

\title{The production of uncertainty in three-dimensional
	Navier-Stokes turbulence}

\author{Jin
	Ge\aff{1}
  \corresp{\email{jin.ge@centrale.centralelille.fr}}, Joran
  Rolland\aff{1}\corresp{\email{joran.rolland@centralelille.fr}} \and John
  Christos
  Vassilicos\aff{1}\corresp{\email{john-christos.vassilicos@centralelille.fr}}}

\affiliation{\aff{1}Univ. Lille, CNRS, ONERA, Arts et Métiers ParisTech,
	Centrale Lille, UMR 9014 - LMFL - Laboratoire de Mécanique des
	Fluides de Lille - Kampé de Feriet, F-59000 Lille, France}

\begin{document}
\maketitle

\begin{abstract}
We derive the evolution equation of the average uncertainty energy for
periodic/homogeneous incompressible Navier-Stokes turbulence and show
that uncertainty is increased by strain rate compression and decreased
by strain rate stretching. We use three different direct numerical
simulations (DNS) of non-decaying periodic turbulence and identify a
similarity regime where (a) the production and dissipation rates of
uncertainty grow together in time, (b) the parts of the uncertainty
production rate accountable to average strain rate properties on the 
one hand and fluctuating strain rate properties on the other also grow
together in time, (c) the average uncertainty energies along the three
different strain rate principal axes remain constant as a ratio of the
total average uncertainty energy, (d) the uncertainty energy
spectrum's evolution is self-similar if normalised by the
uncertainty's average uncertainty energy and characteristic length and
(e) the uncertainty production rate is extremely intermittent and
skewed towards extreme compression events even though the most likely
uncertainty production rate is zero. Properties (a), (b) and (c) imply
that the average uncertainty energy grows exponentially in this
similarity time range. The Lyapunov exponent depends on both the
Kolmogorov time scale and the smallest Eulerian time scale, indicating
a dependence on random large-scale sweeping of dissipative eddies.
In the two DNS cases of statistically stationary turbulence, this
exponential growth is followed by an exponential of exponential
growth, which is in turn followed by a linear growth in the one DNS
case where the Navier-Stokes forcing also produces uncertainty.
\end{abstract}


\section{Introduction}   
\label{sec:Introduction}
 It is basic textbook knowledge that turbulent flow realisations are
 not repeatable whereas statistics over many realisations of a
 turbulent flow are \citep{tennekes1972first}. This well-known
 empirical observation suggests the presence of some kind of chaotic
 attractor. The pioneering work of Lorenz has shown the presence of
 chaos and strange attractors and their resulting high sensitivity to
 initial conditions in non-linear systems with a small number of
 degrees of freedom
 \citep{lorenz1963deterministic,sparrow2012lorenz}. \citet{deissler1986navier}
 demonstrated that similar extreme sensitivity to initial conditions
 is also present in fully developed turbulent solutions of the
 Navier-Stokes equation which is a non-linear system with a very large
 number of degrees of freedom, in fact asymptotically infinite with
 increasing Reynolds number. High sensitivity to initial conditions is
 at the root of non-repeatability and therefore
 uncertainty. Uncertainty is present in a wide range of non-linear
 systems with many degrees of freedom such as turbulent flows,
 magnetohydrodynamics \citep{ho2020fluctuations} and plasma physics
 \citep{cheung1987chaotic} and is also at the core of the problem of
 atmospheric predictability
 \citep{lorenz1963deterministic,leith1971atmospheric}. It may not be
 enough, however, to simply rely on the general concepts of chaos and
 strange attractors (and bifurcations) if one wants to understand
 uncertainty. This paper's motivation is to understand uncertainty and
 its growth in the case of Navier-Stokes turbulence in some physically
 concrete terms.
 
 The solutions of the Navier-Stokes equation are velocity and pressure fields which evolve in time. The uncertainty of a time-dependent velocity field $\boldsymbol{u}^{(1)} (\boldsymbol{x},t)$ is measured by its difference from a velocity field $\boldsymbol{u}^{(2)} (\boldsymbol{x},t)$ with near-identical initial conditions: the velocity difference between these two fields at time $t$ is $\Delta \boldsymbol{u}\equiv\boldsymbol{u}^{(2)}-\boldsymbol{u}^{(1)}$. Based on this velocity-difference field, the average uncertainty in the system is measured in terms of its kinetic energy as $\left\langle
 E_{\Delta}\right\rangle\equiv\left\langle\left|\Delta\boldsymbol{u}\right|^{2}/2\right\rangle$, where $\left\langle \cdot \right\rangle$ represents spatial average (over $\boldsymbol{x}$). In the presence of a strange attractor, its chaotic nature is expected to lead to exponential growth of the difference
 between two fields initially very close together \citep{deissler1986navier,ruelle1981small}, i.e.
 \begin{equation}
 \label{eq:exponential growth of uncertainty}
  \frac{{\rm d}~}{{\rm d}t}\left\langle
 E_{\Delta}\right\rangle=2\lambda\left\langle
 E_{\Delta}\right\rangle,
 \end{equation}
 where $\lambda$ is the maximal Lyapunov exponent.
 
 To evaluate the Lyapunov exponent in the case of statistically stationary homogeneous turbulence, \citet{ruelle1979microscopic} argued that when the two fields $\boldsymbol{u}^{(1)}$ and $\boldsymbol{u}^{(2)}$ differ initially only at the very smallest scales, then $\lambda^{-1}$ should be the Kolmogorov time scale $\tau_{\eta}$ i.e. $\lambda^{-1} \sim \tau_{\eta}\equiv \left(\nu/\varepsilon\right)^{1/2}$ where $\nu$ is the fluid's kinematic viscosity and $\varepsilon$ is the turbulence dissipation rate. Kolmogorov equilibrium $\varepsilon \sim U^{3}/L$ for statistically stationary homogeneous turbulence implies $\lambda \sim \tau_{\eta}^{-1} \sim (L/U)^{-1} \text{Re}^{1/2}$ in terms of the large eddy turnover time $L/U$ and the Reynolds number $\text{Re}=UL/\nu$ where $U$ is the rms turbulence velocity and $L$ the integral length scale. Intermittency corrections have been considered in the form $\lambda\sim (L/U)^{-1} \text{Re}^{a}$ where $a=0.459$ instead of $0.5$ (\citet{crisanti1993intermittency} derived this correction on the basis of a multi-fractal model).  Whilst this correction agrees with numerical observations from the shell model \citep{aurell1997predictability}, neither $a=0.459$ nor $a=0.5$ agree with observations from direct numerical simulations (DNS) of Navier-Stokes statistically stationary homogeneous turbulence \citep{berera2018chaotic,boffetta2017chaos}. In fact, the DNS results of \citet{mohan2017scaling} suggest that $\lambda\tau_{\eta}$ increases with Reynolds number, i.e. $a>0.5$, suggesting that time scales smaller than $\tau_{\eta}$ may be at play. Understanding the growth of uncertainty in some physically concrete terms, as stated above, must also involve shedding some light on the scalings of the maximal Lyapunov exponent which clearly remains an
 open question. In fact the question may be even more widely open as a superfast uncertainty growth may have been observed at very early times in some DNS results \citep{li2020superfast}. Such superfast growth is not ruled out by the rigorous constraint on the uncertainty growth derived from the Navier-Stokes equation by \citet{li2013distinction}: $\left\langle
 E_{\Delta}(t)\right\rangle/\left\langle E_{\Delta}(0)\right\rangle\leq\exp(\sigma\sqrt{\text{Re}}\sqrt{t}+\sigma_{1}t)$ where $\sigma$ and $\sigma_{1}$ are the coefficients depending on the perturbations.
 
 The difference between the velocity fields $\boldsymbol{u}^{(1)}$ and $\boldsymbol{u}^{(2)}$ may be expected to grow in a way that develops differences over length scales $l$ larger than the very smallest scales. When this happens, one may assume equation (\ref{eq:exponential growth of uncertainty}) to remain valid but with a maximal Lyapunov exponent which reflects the characteristic time at length scale $l$, i.e. $\lambda^{-1} \sim\tau_{l}\equiv E_{l}/\varepsilon$ where $E_l$
 is the kinetic energy characterising length scale $l$ \citep{lorenz1969predictability}. It may then be natural to expect $\left\langle E_{\Delta}\right\rangle\sim E_l$ \cite{aurell1997predictability} which leads to a linear growth of $\left\langle E_{\Delta}\right\rangle$ from equation (\ref{eq:exponential growth of uncertainty}) and $\lambda^{-1} \sim E_{l}/\varepsilon$. A linear growth has been widely reported in numerical experiments using the Eddy Damped Quasi-Normal Markovian (EDQNM) closure \citep{leith1972predictability}, shell models \citep{aurell1997predictability} and DNS \citep{berera2018chaotic,boffetta2017chaos}.
 
 There have already been some attempts at understanding uncertainty in physically concrete terms. \citet{boffetta1997predictability} investigated the growth of uncertainty in two-dimensional decaying homogeneous turbulence and found that the uncertainty growth is ruled by the error located in the positions of vortices. \citet{mohan2017scaling} found that much or most of the uncertainty is concentrated near vortex tubes in three-dimensional statistically stationary homogeneous
 turbulence and considered the possibility of local instability mechanisms reminiscent of pairing instabilities of corotating vortices as in mixing layers. \citet{clark2021chaotic,clark2022critical} investigated the dependence of uncertainty on spatial dimension (between 2 to 8) in DNS and in an EDQNM model of statistically stationary homogeneous turbulence. They found a critical dimension $d_{c}\approx5.88$ which is close to the dimension of maximum enstrophy production and above which the turbulence uncertainty is no longer ruled by chaoticity. From these results, \citet{clark2022critical} speculated that vortex stretching and strain self-amplification, which are responsible for enstrophy generation, may also be important for uncertainty generation. The present paper is an effort in the direction of understanding uncertainty growth in terms of vortex stretching and compression dynamics and statistics.
 
 In the following section we derive, from the Navier-Stokes equations, the evolution equation for the uncertainty energy $\left\langle E_{\Delta}\right\rangle$ in the case of periodic/homogeneous
 turbulence. This uncertainty equation involves three different mechanisms: internal production resulting from interactions between the strain rate and the velocity-difference field, dissipation of the velocity-difference field and external force input. We use three different DNS of forced periodic/homogeneous turbulence to study these mechanisms and in section \ref{sec:Numerical steups} we present their numerical setups. Our DNS results and their analysis are presented in section \ref{sec:Numerical results} and we conclude in section \ref{sec:Conclusion}.

\section{Theoretical analysis of the uncertainty }
\label{sec:Theoretical analysis}

In the first part of this section we derive the evolution equation for the uncertainty energy $\left\langle E_{\Delta}\right\rangle$ and in the second part we discuss the production of uncertainty energy.
\subsection{\label{sec:Evolution equation of uncertainty}Evolution equation of uncertainty}
The reference field $\boldsymbol{u}^{(1)}$ and the perturbed field $\boldsymbol{u}^{(2)}=\boldsymbol{u}^{(1)}+\Delta \boldsymbol{u}$ are both governed by the incompressible Navier-Stokes equations
\begin{equation}
\label{eq:NS equation}
\begin{array}{cc}			
\frac{\partial~}{\partial t}u^{(m)}_{i}+u^{(m)}_{j}\frac{\partial~}{\partial x_{j}}u^{(m)}_{i}=-\frac{\partial~}{\partial x_{i}}p^{(m)}+\nu\frac{\partial^{2}~}{\partial x_{j}\partial x_{j}}u^{(m)}_{i}+f^{(m)}_{i},\\[6pt]
\frac{\partial~}{\partial x_{i}}u^{(m)}_{i}=0,
\end{array}
\end{equation}
where $p$ is the pressure to density ratio, $\boldsymbol{f} = (f_1, f_2, f_3)$ is the force per unit mass field, and the number $m=1$ or $2$ in the superscript parentheses indicates whether the velocity/pressure field is the reference or the perturbed one. The equation for $\Delta\boldsymbol{u}\equiv\boldsymbol{u}^{(2)}-\boldsymbol{u}^{(1)}$ follows and is 
	\begin{equation}
\label{eq:difference NS equation}
\begin{array}{cc}	
\frac{\partial~}{\partial t}\Delta
u_{i}+u^{(1)}_{j}\frac{\partial~}{\partial
	x_{j}}\Delta u_{i}+\Delta
u_{j}\frac{\partial~}{\partial x_{j}}\Delta
u_{i}+\Delta u_{j}\frac{\partial~}{\partial
	x_{j}}u^{(1)}_{i}=-\frac{\partial~}{\partial
	x_{i}}\Delta p+\nu\frac{\partial^{2}~}{\partial
	x_{j}\partial x_{j}}\Delta u_{i}+\Delta
f_{i},\\[6pt] \frac{\partial~}{\partial x_{i}}\Delta
u_{i}=0,
\end{array}
\end{equation}
where $\Delta p \equiv p^{(2)}-p^{(1)}$ and $\Delta\boldsymbol{f} \equiv\boldsymbol{f}^{(2)}-\boldsymbol{f}^{(1)}$ are the pressure and forcing differences respectively. The divergence-free property of $\boldsymbol{u}^{(m)}$ implies that $\Delta\boldsymbol{u}$ is also divergence-free. Multiplying both sides of equation (\ref{eq:difference NS equation}) with $\Delta u_{i}$, summing over $i=1,2,3$ and using incompressibility we obtain
\begin{eqnarray}
\label{eq:singlepoint uncertainty equation}
\frac{\partial~}{\partial t}E_{\Delta}+\frac{\partial~}{\partial x_{j}}\left(u^{(1)}_{j} E_{\Delta}\right)&+&\frac{\partial~}{\partial x_{j}}\left(\Delta u_{j}E_{\Delta}\right)+\Delta u_{i}\Delta u_{j}\frac{\partial~}{\partial x_{j}}u^{(1)}_{i}=\nonumber\\
&-&\frac{\partial~}{\partial x_{i}}\left( \Delta u_{i}\Delta p\right)+\nu\frac{\partial~}{\partial x_{j}}\left(\frac{\partial E_{\Delta}}{\partial x_{j}}\right)-\nu\frac{\partial \Delta u_{i}}{\partial x_{j}}\frac{\partial \Delta u_{i}}{\partial x_{j}}+\Delta f_{i}\Delta u_{i}.
\end{eqnarray}
The second and third terms on the left-hand side of equation (\ref{eq:singlepoint uncertainty equation}), as well as the first and second terms on the right-hand side, are in flux form. In the case of periodic/homogeneous turbulence, these four terms average to zero when a spatial average is applied to them, and therefore equation (\ref{eq:singlepoint uncertainty equation}) leads to 
\begin{equation}
\label{eq:total uncertainty equation}
\frac{{\rm d}~}{{\rm d} t}\left\langle E_{\Delta}\right\rangle
=\left\langle
P_{\Delta}\right\rangle-\left\langle\varepsilon_{\Delta}\right\rangle+\left\langle
F_{\Delta}\right\rangle,
\end{equation}
where

\refstepcounter{equation}
$$
P_{\Delta}=-\Delta u_{i}S_{ij}^{(1)}\Delta u_{j}, \label{eq:production}\quad
\varepsilon_{\Delta}=\nu\frac{\partial \Delta u_{i}}{\partial x_{j}}\frac{\partial \Delta u_{i}}{\partial x_{j}}, \quad
F_{\Delta}=\Delta f_{i}\Delta u_{i}\eqno{(\theequation{\mathit{a},\mathit{b},\mathit{c}})}\label{P epsilon F}
$$
and $S_{ij}^{(1)}=\left(\partial u_{i}^{(1)}/\partial x_{j}+\partial u_{j}^{(1)}/\partial x_{i}\right)/2$ is the reference field's strain rate tensor. 

In periodic/homogeneous turbulence the average uncertainty energy evolves via (i) dissipation of uncertainty which always reduces uncertainty because the dissipation rate $\varepsilon_{\Delta}$ is always positive, (ii) external input/output of uncertainty with rate $F_{\Delta}$ which depends on the
force-difference field $\Delta \boldsymbol{f}$, and (iii) internal production of uncertainty via the production rate $P_{\Delta}$. In the absence of external force difference (i.e. $\Delta \boldsymbol{f} = 0$), uncertainty can only grow because of internal production in which case $\left\langle P_{\Delta}\right\rangle$ should be positive and greater than $\left\langle \varepsilon_{\Delta}\right\rangle$.

Note that both fields $\boldsymbol{u}^{(1)}$ and $\boldsymbol{u}^{(2)}$ can be taken as the reference field and we therefore must have $\left\langle P_{\Delta}\right\rangle=-\left\langle\Delta u_{i}S_{ij}^{(1)}\Delta u_{j}\right\rangle=-\left\langle\Delta u_{i}S_{ij}^{(2)}\Delta u_{j}\right\rangle$ in periodic/homogeneous turbulence. Indeed, defining $\Delta S_{ij}=\left(\partial \Delta u_{i}/\partial x_{j}+\partial \Delta u_{j}/\partial x_{i}\right)/2$, we have $S^{(2)}_{ij}=S^{(1)}_{ij}+\Delta S_{ij}$ and $\left\langle P_{\Delta}\right\rangle=-\left\langle\Delta u_{i}S_{ij}^{(1)}\Delta u_{j}\right\rangle=-\left\langle\Delta u_{i}S_{ij}^{(2)}\Delta u_{j}\right\rangle-\left\langle\Delta u_{i}\Delta S_{ij}\Delta u_{j}\right\rangle$. Given that
$\Delta\boldsymbol{u}$ is divergence-free, we also have $\Delta u_{i}\Delta S_{ij}\Delta u_{j}=\frac{1}{2}\left(\frac{\partial~}{\partial x_{j}}\left(\Delta u_{j}E_{\Delta}\right)+\frac{\partial~}{\partial x_{i}}\left(\Delta u_{i}E_{\Delta}\right)\right)$ which implies $\left\langle\Delta u_{i}\Delta S_{ij}\Delta u_{j}\right\rangle=0$ for periodic/homogeneous turbulence. Hence, $\left\langle P_{\Delta}\right\rangle=-\left\langle\Delta u_{i}S_{ij}^{(1)}\Delta u_{j}\right\rangle=-\left\langle\Delta u_{i}S_{ij}^{(2)}\Delta
u_{j}\right\rangle$.

\subsection{\label{sec:Production term}Production of uncertainty}
To consolidate the interpretation of $P_{\Delta}$ as internal production rate of uncertainty, we write
\begin{equation}
\label{eq:correlqtion and decorrelation}
E_{\Delta} + E_{\text{corr}}= E_{\text{tot}}
\end{equation}
where
$E_{\text{tot}}=E^{(1)}+E^{(2)}=\left(\left|\boldsymbol{u}^{(1)}\right|^{2}+\left|\boldsymbol{u}^{(2)}\right|^{2}\right)/2$
and
$E_{\text{corr}}=\boldsymbol{u}^{(1)}\cdot\boldsymbol{u}^{(2)}$. $\left\langle
E_{\text{tot}}\right\rangle$ represents the average total kinetic energy of
the reference and the perturbed velocity fields. Its rate of
change follows from equation (\ref{eq:NS equation}) and is
\begin{equation}
\label{eq:system energy}
\frac{{\rm d}~}{{\rm d} t}\left\langle
E_{\text{tot}}\right\rangle=-\sum_{m=1}^{2}\left\langle\varepsilon^{(m)}\right\rangle+\sum_{m=1}^{2}\left\langle
F^{(m)}\right\rangle,
\end{equation} 
where

\refstepcounter{equation}
$$
\varepsilon^{(m)}\equiv\nu\frac{\partial u_{i}^{(m)}}{\partial x_{j}}\frac{\partial u_{i}^{(m)}}{\partial x_{j}}, \quad
F^{(m)}\equiv f_{i}^{(m)}u_{i}^{(m)}.\eqno{(\theequation{\mathit{a},\mathit{b}})}
$$
If the two velocity fields $\boldsymbol{u}^{(1)}$ and $\boldsymbol{u}^{(2)}$ are so perfectly correlated that they are identical, then $E_{\text{corr}} = E_{\text{tot}}$ and $E_{\Delta}=0$. If, however, these two velocity
fields are totally uncorrelated, then $\left\langle E_{\text{corr}}\right\rangle = 0$ and $\left\langle
E_{\Delta}\right\rangle = \left\langle E_{\text{tot}}\right\rangle$. The average internal production rate of uncertainty $\left\langle P_{\Delta}\right\rangle$ is an internal transfer rate between $\left\langle E_{\text{corr}}\right\rangle$ and $\left\langle E_{\Delta}\right\rangle$, i.e. a transfer rate from correlation to decorrelation if it is positive and from decorrelation to correlation if it is negative. Indeed, from equations (\ref{eq:correlqtion and decorrelation}), (\ref{eq:system energy}) and (\ref{eq:total uncertainty equation}), we have
\begin{equation}
\label{eq:total correlation equation}
\frac{{\rm d~}}{{\rm d} t}\left\langle
E_{\text{corr}}\right\rangle=-\left\langle
P_{\Delta}\right\rangle-\left\langle\varepsilon_{\text{corr}}\right\rangle+\left\langle
F_{\text{corr}}\right\rangle,
\end{equation}
where

\refstepcounter{equation}
$$
\varepsilon_{\text{corr}}=\nu\frac{\partial u_{i}^{(1)}}{\partial x_{j}}\frac{\partial u_{i}^{(2)}}{\partial x_{j}}, \quad
F_{\text{corr}}=f_{i}^{(1)}u_{i}^{(2)}+f_{i}^{(2)}u_{i}^{(1)}.\eqno{(\theequation{\mathit{a},\mathit{b}})}
$$
so that $\left\langle P_{\Delta}\right\rangle$ appears with opposite
signs in equation (\ref{eq:total uncertainty equation}) and in equation (\ref{eq:total correlation equation}) and is absent from equation (\ref{eq:system energy}). If the two flows are identical, i.e. $\boldsymbol{u}^{(1)}=\boldsymbol{u}^{(2)}$, then $P_{\Delta}=\varepsilon_{\Delta}=F_{\Delta}=0$, and if they are totally uncorrelated, then $\left\langle P_{\Delta} \right\rangle = \left\langle \varepsilon_{\text{corr}} \right\rangle = \left\langle F_{\text{corr}} \right\rangle = 0$.

According to equation (\ref{eq:total uncertainty equation}), the evolution of the average uncertainty energy depends on the reference field via its strain rate tensor in the uncertainty production term. The incompressible Navier-Stokes evolution of the strain rate tensor is given by
  \begin{equation}
\label{eq:evolution of strain rate}
\frac{\partial}{\partial t}S_{ij}+u_{k}\frac{\partial}{\partial x_{k}}S_{ij}=-S_{ik}S_{kj}-
\frac{1}{4}\left(\omega_{i}\omega_{j}-\delta_{ij}\left|\boldsymbol{\omega}\right|^{2}\right)-P_{ij}+\nu\frac{\partial^{2}}{\partial x_{j}\partial x_{j}}S_{ij}+F_{ij},
\end{equation} 	
where $\boldsymbol{\omega} \equiv \nabla\times\boldsymbol{u}$ is the vorticity, $\delta_{ij}$ is the Kronecker delta, $P_{ij} \equiv \partial^{2}p/\partial x_{i}\partial x_{j}$ is the pressure Hessian tensor and $F_{ij} \equiv (\partial f_{i}/\partial x_{j}+\partial f_{j}/\partial x_{i})/2$. The first and second terms in the right-hand side of equation (\ref{eq:evolution of strain rate}) represent strain self-amplification and vortex-stretching respectively. They enhance the flow's strain rate once and where it is non-negligibly present, while the pressure Hessian induces its initial growth where it is negligibly small but contributes less to its further development \citep{paul2017genesis}. Therefore, the internal production of uncertainty can be related to the strain self-amplification and
vortex-stretching as speculated by \citet{clark2022critical} in their conclusion, but also to the pressure Hessian. $F_{ij}$ in equation (\ref{eq:evolution of strain rate}) represents the influence of the external forcing on the strain rate tensor. If the external forcing and its spatial gradients are not zero but there is no force difference in the system, i.e., $\Delta\boldsymbol{f}=0$
and therefore $F_{\Delta}=0$, then there is no direct external generation or depletion of uncertainty in equation (\ref{eq:total uncertainty equation}) but the external forcing does nevertheless influence the strain rate tensor's evolution because of $F_{ij}$ in equation (\ref{eq:evolution of strain rate}) and thereby indirectly influences the evolution of the internal production of uncertainty in equation (\ref{eq:total uncertainty equation}).

The presence of the strain rate tensor in the internal uncertainty production reveals the critical and opposing roles of compression and stretching motions in the generation and reduction of
uncertainty. Using the principal axes of $S^{(1)}_{ij}$ (or $S^{(2)}_{ij}$) as a local orthonormal reference frame, we can write
\begin{equation}
\label{eq:Production in principal axe}
P_{\Delta}=-\left(\Lambda^{(1)}_{1}\Delta w_{1}^{2}+\Lambda^{(1)}_{2}\Delta w_{2}^{2}+\Lambda^{(1)}_{3}\Delta w_{3}^{2}\right) ,
\end{equation} 	   
where $\Lambda^{(1)}_{1}$, $\Lambda^{(1)}_{2}$, $\Lambda^{(1)}_{3}$ are the eigenvalues of $S^{(1)}_{ij}$ and $\Delta w_{1}$, $\Delta w_{2}$, $\Delta w_{3}$ are the components of the velocity-difference vector projected on the
corresponding principal axes. Incompressibility forces $S_{ij}^{(1)}$ to be traceless, i.e., $\Lambda^{(1)}_{1}+\Lambda^{(1)}_{2}+\Lambda^{(1)}_{3}=0$. Defining the order of eigenvalues as $\Lambda^{(1)}_{1}\leq\Lambda^{(1)}_{2}\leq\Lambda^{(1)}_{3}$, we must have $\Lambda^{(1)}_{1}<0$ representing local compression and $\Lambda^{(1)}_{3}>0$ representing local stretching \citep{ashurst1987alignment}, while the
sign of intermediate eigenvalue is uncertain but has been found to most likely be positive in DNS of turbulent flows \citep{ashurst1987alignment}. The important point which can now be made on the basis of equation (\ref{eq:Production in principal axe}) is that uncertainty is always produced in the compressive direction ($\Lambda^{(1)}_{1}<0$) and always attenuated in the stretching direction ($\Lambda^{(1)}_{3}>0$). In the absence of external input of uncertainty, the growth of average uncertainty energy can only occur through compression events, and only if compression overwhelms stretching in $\left\langle P_{\Delta} \right\rangle$ and determines its sign. Spontaneous decorrelation of a flow from its perturbed flow in the absence of external inputs of uncertainty can only occur through local compressions.

\section{\label{sec:Numerical steups}Numerical setups}
To study the growth of average uncertainty energy in
periodic/homogeneous turbulence, we use a fully de-aliased
pseudo-spectral code to perform DNS of forced incompressible
Navier-Stokes turbulence in a periodic box of size
$\mathcal{L}^{3}=(2\pi)^{3}$. Time advancement is achieved with a
second-order Runge-Kutta scheme. The code strategy is detailed by
\citet{vincent1991spatial}. In all our
  simulations, the number of grid points is $N^{3}=512^{3}$ and the
  spatial resolution $\left\langle k_{\max}\eta\right\rangle_{t}$ (see
  definition in caption of Table 1) is between 1.6 and 1.7. The time
step is calculated by the CFL condition and the CFL number is $0.4$.
We first generate a reference field and copy it but generate randomly the velocity field in the perturbed wavenumber range to create the perturbed flow at a time which we refer to as $t_{0}=0$,
  i.e. $\boldsymbol{u}^{(2)}(\boldsymbol{x},t_{0})$. In Fourier space, $\hat{\boldsymbol{u}}^{(2)}(\boldsymbol{k},t_0)$ in each wavevector has six components:
  	\begin{equation}
  	\label{eq:initial perturbations}
  	\hat{\boldsymbol{u}}^{(2)}(\boldsymbol{k},t_{0})=\left(
  	\begin{array}{ccc}
  		 u_{x_{0}}^{(2)}(\boldsymbol{k})e^{i\phi_{x_{0}}(\boldsymbol{k})}\\
  		 u_{y_{0}}^{(2)}(\boldsymbol{k})e^{i\phi_{y_{0}}(\boldsymbol{k})}\\
  		 u_{z_{0}}^{(2)}(\boldsymbol{k})e^{i\phi_{z_{0}}(\boldsymbol{k})}\\
  	\end{array}\right),
  \end{equation} 
  which follow three constraints
  \begin{enumerate}
  	\item Incompressibility :
  	\begin{equation}
  		\imath\boldsymbol{k}\cdot\hat{\boldsymbol{u}}^{(2)}(\boldsymbol{k},t_{0})=0\,.	\label{const_inc}
  	\end{equation}
  	\item The initial energy spectra of the reference flow and the perturbed flow are identical:
  	\begin{equation}			
  		\hat{E}^{(1)}(k,t_{0})=\int_{\left|\boldsymbol{k}\right|=k}\frac{\left|\hat{\boldsymbol{u}}^{(1)}(\boldsymbol{k},t_{0})\right|^{2}}{2}\mathrm{d}^{2}\boldsymbol{k}
  		 =\int_{\left|\boldsymbol{k}\right|=k}\frac{\left|\hat{\boldsymbol{u}}^{(2)}(\boldsymbol{k},t_{0})\right|^{2}}{2}\mathrm{d}^{2}\boldsymbol{k}=\hat{E}^{(2)}(k,t_{0}).
  		\label{const_E}
  	\end{equation}
  	\item The difference initially only exists in
  	the smallest scales, i.e. $\left|\boldsymbol{k}\right|>k_{0}$ where
  	$k_{0}=0.9k_{max}$ and $k_{max} = N/3$ is the maximum resolvable wavenumber after
  	de-aliasing (see however Appendix \ref{app:Sensitivity of the
  		uncertainty energy to the initial perturbation} for
  	different perturbed wavenumber ranges):
  	\begin{equation}
  		\hat{\boldsymbol{u}}^{(2)}(\boldsymbol{k},t_0)=\left\{
  		\begin{array}{ccc}
  			\hat{\boldsymbol{u}}^{(1)}(\boldsymbol{k},t_0)&       &\text{if } \left|\boldsymbol{k}\right|< k_{0}, \\
  			\text{Randomly generated}&       &\text{if } \left|\boldsymbol{k}\right|\geq k_{0}.
  		\end{array}\right.\label{const_dec}
  	\end{equation}
  \end{enumerate}
  For the generation of $\hat{\boldsymbol{u}}^{(2)}(\boldsymbol{k},t_0)$ in the perturbed wavnumber range, these three constraints \emph{a priori} couple all the
  $\hat{\boldsymbol{u}}^{(2)}(\boldsymbol{k},t_0)$ on the sphere of Fourier
  space such that $|\boldsymbol{k}|=k$.  For simplicity of implementation,
  we use a version of equation~(\ref{const_E}) restricted to each
  $\boldsymbol{k}$, such that the sum of the resulting
  $\hat{\boldsymbol{u}}^{(2)}(\boldsymbol{k},t_0)$ over $|\boldsymbol{k}|=k$
  verifies equation~(\ref{const_E}). This means that for each wavevector
  $\boldsymbol{k}$ we compute six random values, three moduli and three
  phases $\left[u_{x_{0}}^{(2)}(\boldsymbol{k}),u_{y_{0}}^{(2)}(\boldsymbol{k}),u_{z_{0}}^{(2)}(\boldsymbol{k}),\phi_{x_{0}}(\boldsymbol{k}), \phi_{y_{0}}(\boldsymbol{k}),\phi_{z_{0}}(\boldsymbol{k})\right] \in[0,\sqrt{2\hat{E}^{(1)}}(k,t_{0})]^3 \times[0,2\pi)^3$
  that follow two constraints coming from the real and imaginary part of
  the imcompressibility condition (equation~(\ref{const_inc})) and one
  constraint from the spectrum (equation~(\ref{const_E})).
  This means that only three independent components have to be drawn and
  the three others will follow.  In practice:
  \begin{itemize}
  	\item In the general case of $k_x\ne 0$, $k_y\ne 0$ and $k_z\ne 0$, two uniform random numbers are drawn in $[0,1)$ yielding $\phi_{x_{0}}(\boldsymbol{k})$ and $\phi_{y_{0}}(\boldsymbol{k})$ after rescaling and
  	one uniform random number in $[0,1)$ yielding $u_{x_{0}}^{(2)}(\boldsymbol{k})$ after rescaling.  The moduli $u_{y_{0}}^{(2)}(\boldsymbol{k})$ and $u_{z_{0}}^{(2)}(\boldsymbol{k})$ are
  	successively computed using equation~(\ref{const_E}).  The sine and
  	the cosine of the phase $\phi_{z_{0}}(\boldsymbol{k})$ are finally
  	computed respectively using the real and imaginary part of the
  	incompressibility condition (equation~(\ref{const_inc})).
  	\item In the case where only one component of the wavevector is equal to zero: 
  	the modulus and the phase in the direction of the zero component of
  	the wavevector are drawn first uniformly from
  	$[0,\sqrt{2\hat{E}^{(1)}}(k,t_{0})]\times[0,2\pi)$. The two other moduli are
  	computed using (equation~(\ref{const_E})), one phase is drawn from
  	$[0,2\pi)$ and the other is deduced from incompressibility.
  	\item In the case where two components of the wavevector are equal to zero: 
  	the real and imaginary parts of incompressibility impose that the
  	modulus of the corresponding component of $\hat{\boldsymbol{u}}^{(2)}$ is
  	zero, and that the corresponding phase is irrelevant.  As a
  	consequence, out of the four remaining values to be determined, one is
  	constrained by equation~(\ref{const_E}).  In practice the two
  	remaining phases are drawn uniformly in $[0,2\pi)^2$, one modulus is
  	drawn uniformly in $[0,\sqrt{2\hat{E}^{(1)}}(k,t_{0})]$ and the other is
  	determined using (equation~(\ref{const_E})).  \end{itemize}

In this way, the initial perturbations, defined as $\Delta\boldsymbol{u}(\boldsymbol{x},t_{0})=\boldsymbol{u}^{(2)}(\boldsymbol{x},t_{0})-\boldsymbol{u}^{(1)}(\boldsymbol{x},t_{0})$, are also incompressible and exist only in the perturbed wavenumber range. Furthermore, the perturbed flow is generated randomly in its perturbed wavenumber range, hence the reference flow and the perturbed flow are initially completely decorrelated in this wavenumber range, which implies.
\begin{equation}
	\hat{E}_{\Delta}(k,t_{0})=\int_{\left|\boldsymbol{k}\right|=k}\frac{\left|\Delta\hat{\boldsymbol{u}}(\boldsymbol{k},t_{0})\right|^{2}}{2}\mathrm{d}^{2}\boldsymbol{k}=\left\{
	\begin{array}{ccc}
		0&       &\text{if } \left|\boldsymbol{k}\right|< k_{0}, \\
		\hat{E}_{\text{tot}}(k,t_{0})&       &\text{if } \left|\boldsymbol{k}\right|\geq k_{0},
	\end{array}\right.
\end{equation}
where $\hat{E}_{\text{tot}}(k,t_{0})=\hat{E}^{(1)}(k,t_{0})+\hat{E}^{(2)}(k,t_{0})$.

Three different cases (F1, F2 and F3) are simulated by applying
different external forcings and initial conditions. In the first case,
labelled F1, a negative damping forcing is applied to both the
reference and the perturbed turbulent fields and the force-difference
field does not vanish. The forcing function is divergence-free as it
depends on the low wavenumber modes of the velocity in Fourier space
as follows
  \begin{equation}
\label{eq:negative damping forcing 1}
\hat{\boldsymbol{f}}^{(m)}\left(\boldsymbol{k},t\right)=\left\{
\begin{array}{ccc}
\frac{\varepsilon_{0}}{2E_{f}^{(m)}}\hat{\boldsymbol{u}}^{(m)}\left(\boldsymbol{k},t\right)&       &\text{if } 0<\left|\boldsymbol{k}\right|\leq k_{f},\\
0&       &\text{otherwise,}
\end{array}\right.
\end{equation}
where $\hat{\boldsymbol{f}}$ and $\hat{\boldsymbol{u}}$ are the
Fourier transforms of $\boldsymbol{f}$ and $\boldsymbol{u}$
respectively, $\varepsilon_{0}$ is the preset average turbulence
dissipation rate and $E_{f}$ is the kinetic energy contained in the
forcing bandwidth $0<\left|\boldsymbol{k}\right|\leq k_{f}$. This
forcing has been widely used to simulate statistically steady
homogeneous isotropic turbulence (HIT) on the computer
\citep{ho2020fluctuations,berera2018chaotic,boffetta2017chaos,mohan2017scaling,clark2022critical,clark2021chaotic}. It
offers the advantage of setting the average turbulence dissipation a
priori for statistically steady turbulence. In the present work, we
set $\varepsilon_{0}=0.1$ and $k_{f}=2.5$.

To generate the reference flow we use a von K\'arm\'an initial energy
spectrum with the same coefficients as
\citet{https://doi.org/10.48550/arxiv.1306.3408} and random initial
Fourier phases. We integrate the reference flow till it reaches a
statistically steady state and then seed it with perturbations to
create the perturbed flow at a time which we refer to as
$t_{0}=0$. One can see from equation (\ref{eq:negative damping forcing
  1}) that the external forcings are determined separately by the two
fields and therefore $\Delta \boldsymbol{f} \neq 0$. F1 is the
only one of our three cases where $F_{\Delta}$ is not identically zero
and some uncertainty is introduced by the forcing in equation
(\ref{eq:total uncertainty equation}).

The case F2 is identical to F1 except for the external forcing which
is such that $F_{\Delta}=0$. The forcing in the perturbed field is
determined by the velocity in the reference field as
\begin{equation}
\label{eq:negative damping forcing 2}
\hat{\boldsymbol{f}}^{(2)}\left(\boldsymbol{k},t\right)=\hat{\boldsymbol{f}}^{(1)}\left(\boldsymbol{k},t\right)=\left\{
\begin{array}{ccc}
\frac{\varepsilon_{0}}{2E_{f}^{(1)}}\hat{\boldsymbol{u}}^{(1)}\left(\boldsymbol{k},t\right)&       &\text{if } 0<\left|\boldsymbol{k}\right|\leq k_{f},\\
0&       &\text{otherwise,}
\end{array}\right.
\end{equation}		 
where $\varepsilon_{0}=0.1$ and $k_{f}=2.5$. Therefore, there is no forcing difference between the two fields and all the uncertainty in equation (\ref{eq:total uncertainty equation}) is generated exclusively by the internal production.



The case F3 differs in one essential way from F1
  and F2: rather than force the turbulence into a stationary steady
  state and then introduce the uncertainty after stationarity has set
  in (as in F1 and F2), in F3 we introduce the uncertainty well before
  stationarity has set in, i.e. at a very initial time when the
  initial velocity field has very little energy and the simulation
  starts running with a forcing which eventually brings the turbulence
  into an energetic stationary state. We chose a forcing for F3 that
  is independent of the velocity field to ensure steady buildup of the
  turbulence during a long yet finite time. The initial velocity
fields are randomly generated with the same energy spectrum
$\hat{E}\left(k\right)= 0.3\times10^{-4}k^{-1}$ for the reference and
the perturbed fields and the initial perturbations are seeded in the
high wavenumber Fourier phases in the exact same way as in F1 and
F2. Both flows are forced by an identical single-mode divergence-free
force
\begin{equation}
\label{eq:negative damping forcing 3}
\boldsymbol{f}^{(2)}\left(\boldsymbol{x},t\right)=\boldsymbol{f}^{(1)}\left(\boldsymbol{x},t\right)=\left(
\begin{array}{ccc}
\cos\left(k_{0}y\right)\sin\left(k_{0}z\right)\\
\cos\left(k_{0}z\right)\sin\left(k_{0}x\right)\\
\cos\left(k_{0}x\right)\sin\left(k_{0}y\right)\\
\end{array}\right),
\end{equation}	
where $k_{0}=4$. This forcing differs from F1 but is similar to F2 in
that $F_{\Delta}$ identically vanishes and there is no uncertainty
input from the forcing in equation (\ref{eq:total uncertainty
  equation}). We repeat, however, that the main
  distinguising feature of F3 compared to F1 and F2 is that, in F3,
the reference and the perturbed fields are statistically
non-stationary during their initial growth (driven by the forcing) and
the concurrent initial growth of uncertainty. This non-stationarity
affects equation (\ref{eq:total uncertainty equation}) through the
resulting non-stationarity of the strain rate field in the internal
production rate.

In summary, F1 is the case that is widely used in previous works
\citep{ho2020fluctuations,berera2018chaotic,boffetta2017chaos,mohan2017scaling,clark2022critical,clark2021chaotic}
and F2 differs from it only in terms of $\Delta \boldsymbol{f}$ which
is zero in F2 and non-zero in F1. In both F1 and F2 the perturbation
is made to a fully developed statistically stationary turbulence
whereas in F3 we follow the evolution of two velocity fields which are
initially very weak in terms of energy and very close to each other,
i.e. very highly correlated. Both flows are progressively intensified
by the same spatially sinusoidal time-independent forcing field and
evolve towards statistical stationary fully developed turbulence
while, at the same time, diverging from each
other.

The main parameters characterising the reference flows are given in
table \ref{tab:main parameters} where
  $\left\langle\cdot\right\rangle$ represents the spatial average,
  $\left\langle\cdot\right\rangle_{t}$ represents the temporal average
  and $\left\langle\left\langle\cdot\right\rangle\right\rangle_{t}$
  represents the average in both space and time. For F1 and F2, this
time average is over all time $t\ge 0$ when the reference and
perturbed fields are statistically stationary in the simulations. For
F3, the time average is over the time when the reference flow's
turbulent kinetic energy and dissipation rate are statistically
stationary, i.e. the standard deviations of $\left\langle
E^{(1)}\right\rangle(t)$ and $\left\langle
\varepsilon^{(1)}\right\rangle(t)$ are smaller than $ 8\%$ of
$\left\langle\left\langle E^{(1)}\right\rangle\right\rangle_{t}$ and
$\left\langle\left\langle
\varepsilon^{(1)}\right\rangle\right\rangle_{t}$ respectively. This
leads to $\tau\equiv t/\left\langle T_{0}^{(1)}
\right\rangle_{t}\in[9.4,30.0]$ (Note that the dimensionless time
$\tau \equiv t/\left\langle T_{0}^{(1)} \right\rangle_{t}$ is defined
for all three cases F1, F2 and F3.).
\begin{table}
	\begin{center}
		\def~{\hphantom{0}}
		\begin{tabular}{lccccccccc}
Case&$N^{3}$&$\nu$&$\left\langle\left\langle\varepsilon\right\rangle\right\rangle_{t}$&$\left\langle U\right\rangle_{t}$&$\left\langle L\right\rangle_{t}$&$\left\langle T_{0}\right\rangle_{t}$&$\left\langle\text{Re}\right\rangle_{t}$&$\left\langle\text{Re}_{\lambda}\right\rangle_{t}$&$\left\langle k_{\max}\eta\right\rangle_{t}$\\ [3pt]
\midrule
F1&$512^{3}$&0.0010&0.0981&0.622&1.101&1.771&684.9&151.6&1.70\\
F2&$512^{3}$&0.0010&0.0988&0.622&1.102&1.772&685.4&151.2&1.70\\
F3&$512^{3}$&0.0015&0.4096&0.643&0.345&0.537&148.2&63.8&1.62\\
		\end{tabular}
		\caption{Parameters of the reference flows, where
                  $\left\langle\cdot\right\rangle$ represents the
                  spatial average;
                  $\left\langle\cdot\right\rangle_{t}$ represents the
                  temporal average and
                  $\left\langle\left\langle\cdot\right\rangle\right\rangle_{t}$
                  represents the average in both space and time. $N$
                  is the resolution of the simulations, $\nu$ is the
                  kinematic viscosity, $\varepsilon$ is the
                  dissipation. $U\equiv\sqrt{2\left\langle
                    E\right\rangle/3}$ is the rms velocity and
                  $L\equiv\left(3\pi/4\left\langle
                  E\right\rangle\right)\int k^{-1}\hat{E}(k)\rm{d}$$k$
                  is the integral length scale. $T_{0}\equiv L/U$ is
                  the large eddy turnover time. $\text{Re}\equiv
                  UL/\nu$ is the Reynolds
                  number. $\text{Re}_{\lambda}\equiv Ul_{\lambda}/\nu$
                  is the Reynolds number defined by the Taylor length
                  scale $l_{\lambda}\equiv\sqrt{10\left\langle
                    E\right\rangle\nu/\left\langle\varepsilon\right\rangle}$. $k_{\max}$
                  is the maximum resolvable wavenumber and
                  $\eta\equiv\left(\nu^{3}/\left\langle\varepsilon\right\rangle\right)^{1/4}$
                  is the Kolmogorov scale.}
		\label{tab:main parameters}
	\end{center}
\end{table}
\section{\label{sec:Numerical results}DNS  results}
In this section we present our DNS results concerning equation
(\ref{eq:total uncertainty equation}), starting in subsection
\ref{sec:Time evolution of uncertainty} with the time evolution of
$\left\langle E_{\Delta} \right\rangle$ during the decorrelation
process and an analysis of the three mechanisms at play and of the
uncertainty's energy spectrum. In subsection \ref{sec:Quantitative
  analysis of the uncertainty growth} we relate the growth rate of
$\left\langle E_{\Delta} \right\rangle$ to detailed properties of the
production and dissipation of uncertainty, of the strain rate
eigenvalues and of the distribution of uncertainty energy in the three
principal axes of the strain rate tensor. In particular, we derive the
chaotic exponential growth of $\left\langle E_{\Delta} \right\rangle$
from similarity behaviours of these quantities. In subsection
\ref{sec:The probability distribution of the uncertainty production}
we go beyond the average production of uncertainty and report
probability density functions of $P_{\Delta}$.
\subsection{\label{sec:Time evolution of uncertainty}Time evolution of uncertainty}
\subsubsection{Uncertainty energy}
\begin{figure}
	\centering \subfigure[case F1]{
		\label{fig:time evolution of uncertinty case1}
		\includegraphics[width=0.49\textwidth]{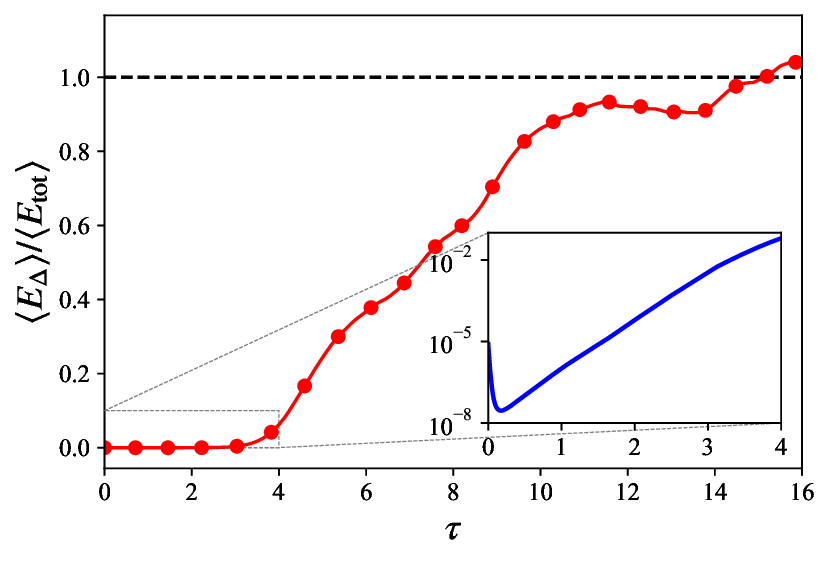}}
	\subfigure[case F2]{
		\label{fig:time evolution of uncertinty case2}
		\includegraphics[width=0.49\textwidth]{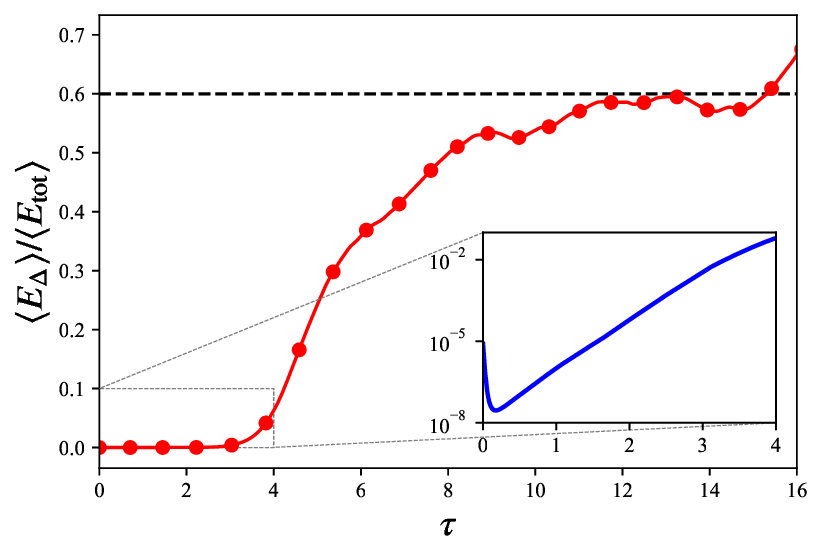}}
	\subfigure[case F3]{
		\label{fig:time evolution of uncertinty case3}
		\includegraphics[width=0.49\textwidth]{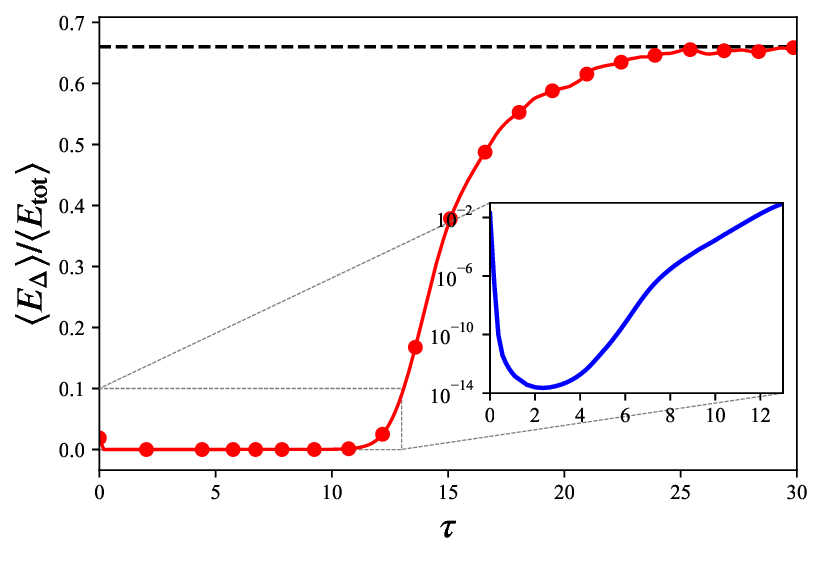}}
	\subfigure[comparison F1 - F2]{
		\label{fig:time evolution of uncertinty compare}
		\includegraphics[width=0.49\textwidth]{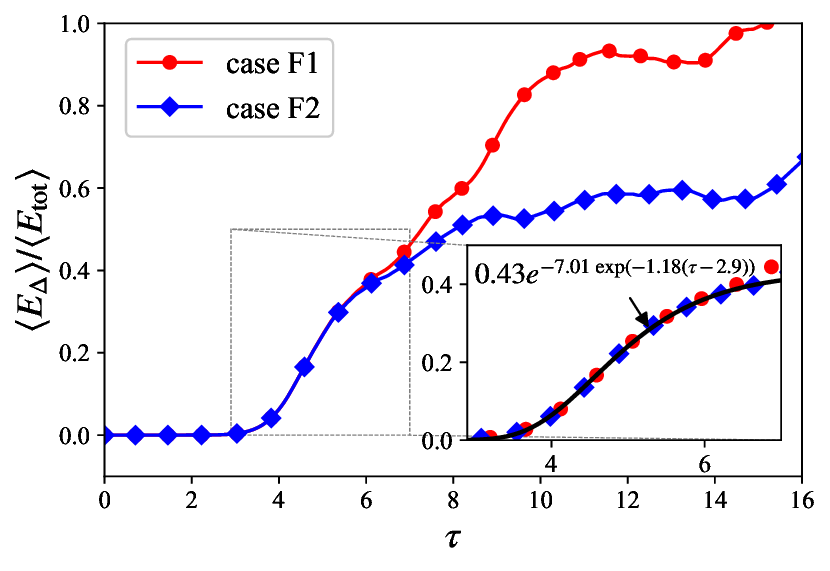}}
	\caption{Time evolutions of average uncertainty energy for
          different cases. Inset: the initial time evolution of
          average uncertainty energy in semilogarithmic plot. The
          uncertainty evolutions of F1 and F2 are presented together
          in (\textit{d}). Inset: the uncertainty evolutions during
          $\tau\in[2.9,6.5]$. The exponential of exponential function
          fit is indicated by a solid black line. }
	\label{fig:time evolution of uncertinty} 
\end{figure}
\begin{figure}
	\centering 
	\subfigure[case F1]{
		\label{fig:uncertainty equation case1}
		\includegraphics[width=0.49\textwidth]{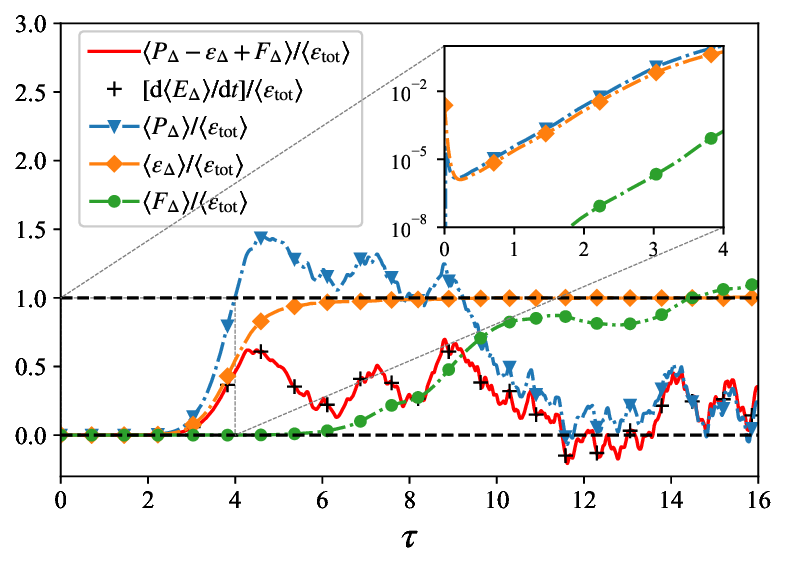}}
	\subfigure[case F2]{
		\label{fig:uncertainty equation case2}
		\includegraphics[width=0.49\textwidth]{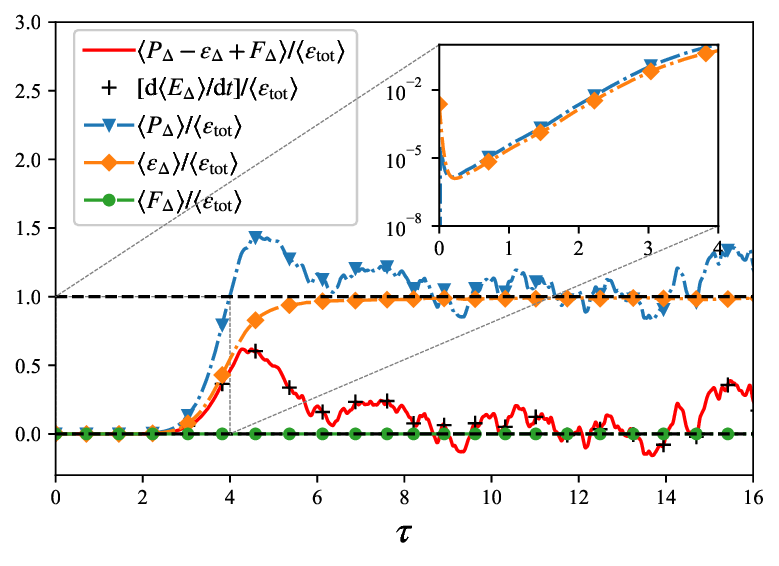}}
	\subfigure[case F3]{
		\label{fig:uncertainty equation case3}
		\includegraphics[width=0.49\textwidth]{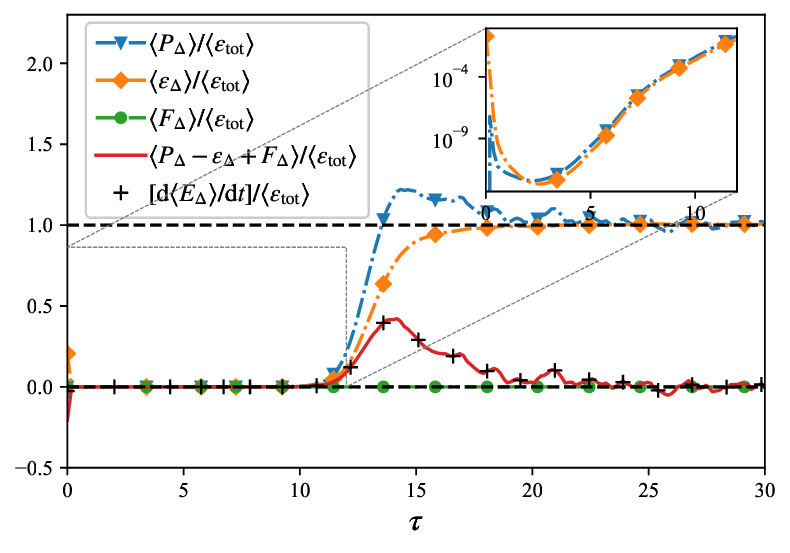}}
	\caption{Time evolutions of each term in equation
          (\ref{eq:total uncertainty equation}) for different cases,
          where
          $\varepsilon_{\text{tot}}=\varepsilon^{(1)}+\varepsilon^{(2)}$
          is the total dissipation of the reference and perturbed
          fields. Inset: the initial time evolution of the internal
          production, dissipation and external input/output in
          semilogarithmic plot.}
	\label{fig:uncertainty equation} 
\end{figure}

Figure \ref{fig:time evolution of uncertinty} shows the time
evolutions of $\left\langle E_{\Delta}\right\rangle$ for each case F1,
F2 and F3. The very first thing that happens immediately after the
perturbations are seeded is a decrease of $\left\langle
E_{\Delta}\right\rangle$ in all three cases. This initial correlating
action is caused by the concentration of the initial perturbations at
the highest wavenumbers where dissipation is high. The insets of
figure \ref{fig:uncertainty equation} show that
$\left\langle\varepsilon_{\Delta} \right\rangle$ is orders of
magnitude higher than $\left\langle P_{\Delta} \right\rangle$ at the
earliest times in all three cases. As time proceeds, the uncertainty's
dissipation rate decreases and its production rate increases till
production overtakes dissipation (see figure \ref{fig:uncertainty
  equation}) and $\left\langle E_{\Delta}\right\rangle$ begins to
grow. This initial growth is shown in the insets of figure
\ref{fig:time evolution of uncertinty} and it differs for F1 and F2 on
the one hand and F3 on the other. For F1 and F2, $\left\langle
E_{\Delta}\right\rangle$ is observed to grow exponentially in the
approximate time-range
$\tau\in[0.2,2.9]$. Previous DNS studies have
already observed such exponential growth
\citep{berera2018chaotic,boffetta2017chaos}. For F3, the initial
growth is from $\tau \approx 2.5$ to $\tau\approx 12.6$ and is
subdivided in two parts. In the time range $\tau\in[2.5,7.5]$, the
turbulence and its strain rate are not statistically stationary and
the time evolution of $\left\langle E_{\Delta}\right\rangle$ is not
exponential. Indeed, the plot of the logarithm of $\left\langle
E_{\Delta}\right\rangle$ versus time in the inset of figure
\ref{fig:time evolution of uncertinty case3} has a positive curvature
in that time range. An exponential growth of $\left\langle
E_{\Delta}\right\rangle$ appears to set in at $\tau\approx 7.5$ and
lasts till about $\tau\approx 12.6$. It is
  noteworthy that an exponential growth of uncertainty also exists in
  F3 and that it starts a little earlier than when stationarity sets
  in. (The exponential regime's time range is longer for F3 than for
  F1 and F2 mainly because of F3's lower Reynolds number as argued in
  Appendix \ref{app:Reynolds-number dependence of the time range of
    the exponential regime}). The results and analysis in the
remainder of this paper confirm these interpretations.

The growths of $\left\langle E_{\Delta}\right\rangle$ are identical in
F1 and F2 (see figure \ref{fig:time evolution of uncertinty compare})
till the time when $\left\langle F_{\Delta}\right\rangle$ becomes
significantly non-zero in F1 (see figure \ref{fig:uncertainty equation
  case1}). The regime of exponential growth is followed by what
appears to be an exponential of exponential regime from
$\tau\approx 2.9$ to
$\tau\approx 6.5$. This exponential of
exponential growth is the same in F1 and F2 and is highlighted by the
fit in the inset of figure \ref{fig:time evolution of uncertinty
  compare}. A similar growth range has been observed in previous DNS
that are similar to F1 and go up to even higher Reynolds numbers
\citep{berera2018chaotic,boffetta2017chaos}. This exponential of
exponential growth is confirmed by our analysis and further DNS
results in subsection \ref{sec:Quantitative analysis of the
  uncertainty growth}.

After time $\tau=6.5$, the uncertainty growths
for F1 and F2 deviate from each other as shown in figure \ref{fig:time
  evolution of uncertinty compare} ($\frac{\left|\left\langle
  E_{\Delta}\right\rangle_{F1}-\left\langle
  E_{\Delta}\right\rangle_{F2}\right|}{\left\langle
  E_{\Delta}\right\rangle_{F1}}>5\%$ and growing as $\tau$ grows above
$6.5$) because $\left\langle
F_{\Delta}\right\rangle$ starts growing significantly above zero
($\left\langle
F_{\Delta}\right\rangle/\left\langle\varepsilon_{\Delta}\right\rangle=0.06$
at $\tau=6.5$) in F1 whereas it is identically
zero in F2 for all time (see figure \ref{fig:uncertainty equation}).
The reference and perturbed fields achieve significant decorrelation
after the exponential growth of $\left\langle E_{\Delta}\right\rangle$
for both F1 and F2, resulting, in case F1, in non-zero values of
$\left\langle F_{\Delta}\right\rangle$ which eventually grow
significantly above the reference field's turbulence dissipation rate,
but only after $\tau=6.5$. The additional
external decorrelating action of the forcing leads to eventually fully
decorrelated reference and perturbed fields in case F1 as the ratio
$\left\langle E_{\Delta}\right\rangle/\left\langle
E_{\text{tot}}\right\rangle$ stops growing and saturates at
$0.97\pm0.07$ after
$\tau=10.6$. In case F2 the identical forcing in
both fields acts as a perpetual partially correlating (rather than
decorrelating) action of the two fields and to a resulting eventual
saturation of $\left\langle E_{\Delta}\right\rangle/\left\langle
E_{\text{tot}}\right\rangle$ at $0.59\pm0.05$ for
$\tau \ge 8.9$ (In Appendix
  \ref{app:Sensitivity of the uncertainty energy to the initial
    perturbation} we provide evidence showing that the early- and
  mid-time evolutions (i.e. the exponential regime and the exponential
  of exponential regime) of the average uncertainty energy are not
  very sensitive to the form and amplitude of the initial
  perturbations.).

For case F3, the growth of $\left\langle E_{\Delta}\right\rangle$ following the exponential regime ending at about $\tau \approx 12.6$ can be seen in figure \ref{fig:time evolution of uncertinty case3} and cannot be fitted by the exponential of exponential function detected in cases F1 and F2 nor any clear linear or power-law growth functions. As in F2, the correlating action of the identical external forcing in both the reference and perturbed fields leads to them remaining partially correlated at all times and to an eventual saturation of $\left\langle E_{\Delta}\right\rangle/\left\langle E_{\text{tot}}\right\rangle$ at $0.66\pm0.01$ for $\tau\ge25.3$.

\begin{table}
	\begin{center}
		\def~{\hphantom{0}}
		\begin{tabular}{lccccccccc}
		Case&Uncertainty regime&Time interval $\tau$\\
		\midrule
		\multirow{5}*{F1}&Initial decrease&$[0,0.2]$\\
		~&Exponential growth&$[0.2,2.9]$\\
		~&Exponential of exponential growth&$[2.9,6.5]$\\
		~&Linear growth&$[6.5,10.6]$\\
		~&Saturation&$[10.6,+\infty]$\\
		\midrule
		\multirow{5}*{F2}&Initial decrease&$[0,0.2]$\\
		~&Exponential growth&$[0.2,2.9]$\\
		~&Exponential of exponential growth&$[2.9,6.5]$\\
		~&Transient growth&$[6.5,8.9]$\\
		~&Saturation&$[8.9,+\infty]$\\
		\midrule
		\multirow{4}*{F3}&Initial decrease&$[0,2.5]$\\
		~&Unsteady initial growth&$[2.5,7.5]$\\
		~&Exponential growth&$[7.5,12.6]$\\
		~&Nonlinear growth&$[12.6,25.3]$\\
		~&Saturation&$[25.3,+\infty]$\\
		\end{tabular}
		\caption{Time ranges of different
                    uncertainty growth regimes.}
		\label{tab:time characteristics}
	\end{center}
\end{table}

We close this subsection by pointing out that the only case of linear
growth that we may have detected in our DNS is for F1 in the time
range $\tau \in [6.5,10.6]$. A linear growth
regime has been predicted by \citet{aurell1997predictability}, however
our simulations suggest that it may depend on the type of
forcing. Furthermore, the Reynolds number of our DNS may not be high
enough to observe it clearly and the very level of Reynolds number
required may itself depend on the external forcing. We examine this
issue again in the following subsections. The time
  ranges of the different uncertainty growth regimes in each case F1,
  F2 and F3 are summarized in table \ref{tab:time characteristics}.
\begin{figure}
	\centering 
	\subfigure[case F1]{
		\label{fig:P-Epsilon case1}
		\includegraphics[width=0.49\textwidth]{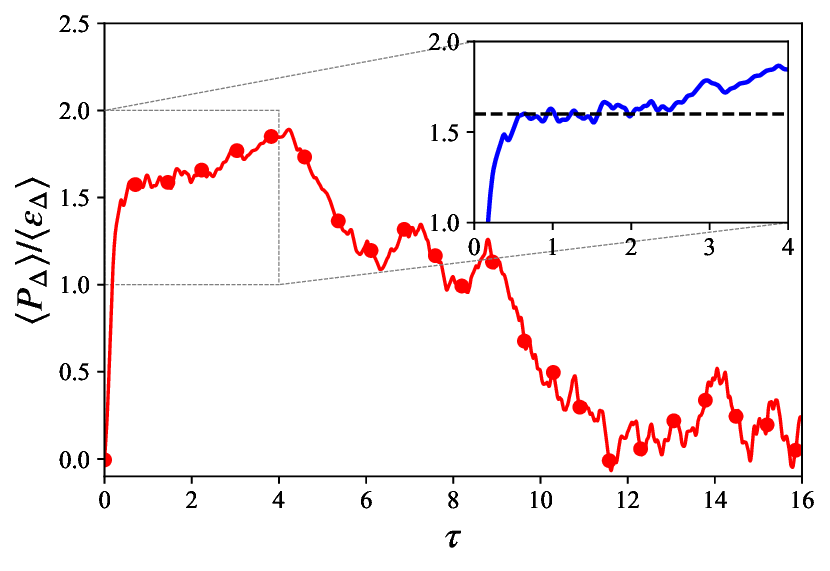}}
	\subfigure[case F2]{
		\label{fig:P-Epsilon case2}
		\includegraphics[width=0.49\textwidth]{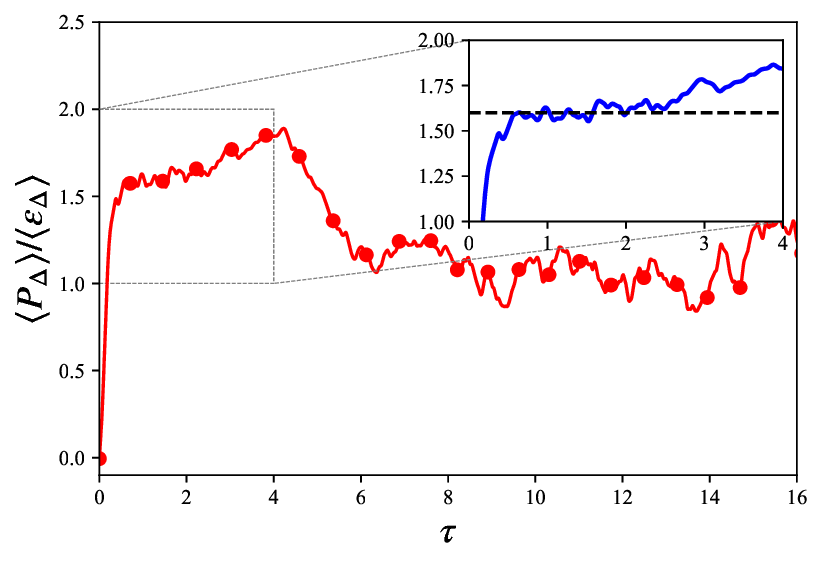}}
	\subfigure[case F3]{
		\label{fig:P-Epsilon case3}
		\includegraphics[width=0.49\textwidth]{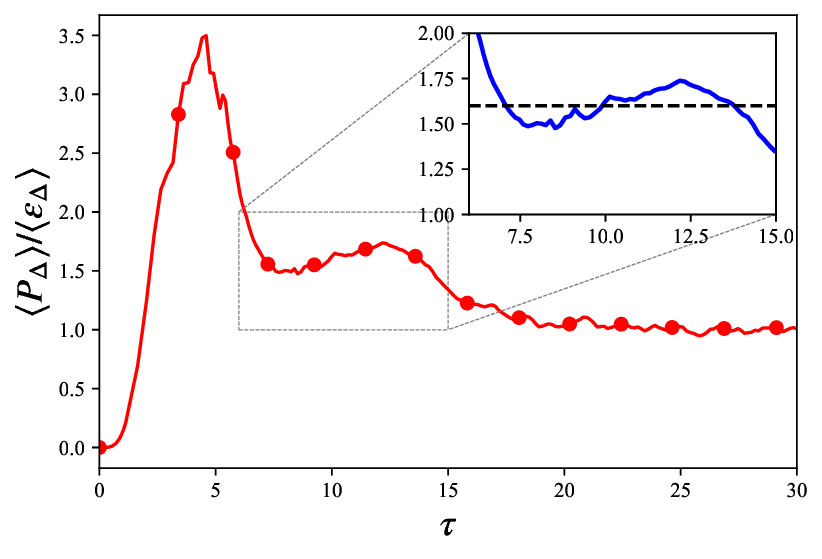}}
	\caption{Time evolutions of ratio $\left\langle P_{\Delta}\right\rangle/\left\langle\varepsilon_{\Delta}\right\rangle$ for different cases. Inset: the evolution of the ratio in the time range of the exponential growth of the average uncertainty.} 
	\label{fig:P-Epsilon} 
\end{figure}

\subsubsection{\label{sec:Mechanisms of the uncertainty evolution}Mechanisms of the uncertainty evolution}
The time evolutions of each term in equation (\ref{eq:total
  uncertainty equation}), including the growth rate ${\rm
  d}\left\langle E_{\Delta}\right\rangle/{\rm d}t$ obtained directly
from the DNS, are shown in figure \ref{fig:uncertainty equation}. As
can be seen in the figure, we started by checking that ${\rm
  d}\left\langle E_{\Delta}\right\rangle/{\rm d}t$ agrees well with
its value obtained from equation (\ref{eq:total uncertainty
  equation}). In all cases F1, F2 and F3, $\left\langle
P_{\Delta}\right\rangle > \left\langle
\varepsilon_{\Delta}\right\rangle$ when ${\rm d}\left\langle
E_{\Delta}\right\rangle/{\rm d}t >0$. In cases F2 and F3 where
$\left\langle F_{\Delta}\right\rangle = 0$ at all times, the eventual
saturation when ${\rm d}\left\langle E_{\Delta}\right\rangle/{\rm d}t
\approx 0$ is characterised by the balance $\left\langle
P_{\Delta}\right\rangle \approx \left\langle
\varepsilon_{\Delta}\right\rangle$. This balance reflects the
long-time partial correlation between the reference and perturbed
fields and the long-time saturation of $\left\langle
E_{\Delta}\right\rangle/\left\langle E_{\text{tot}}\right\rangle$ at a
value smaller than 1 reported in the previous subsection.

We also observe in figure \ref{fig:uncertainty equation} for all cases
F1, F2 and F3 that the long-time saturation is such that
$\left\langle\varepsilon_{\Delta}\right\rangle\approx\left\langle\varepsilon_{\rm
  tot}\right\rangle\equiv\left\langle\varepsilon^{(1)}+\varepsilon^{(2)}\right\rangle$
which implies
$\left\langle\varepsilon_{\text{corr}}\right\rangle\approx0$. In cases
F2 and F3, this means that the long-time saturated non-zero steady
state of $\left\langle P_{\Delta}\right\rangle$ is such that
$\left\langle
P_{\Delta}\right\rangle\approx\left\langle\varepsilon^{(1)}+\varepsilon^{(2)}\right\rangle\approx\left\langle
F^{(1)}+F^{(2)}\right\rangle$ (recall $\left\langle
F_{\Delta}\right\rangle =0$ and $\left\langle F_{\text{corr}}
\right\rangle=0$ in F2, F3): the correlating action by the identical
forcing in both statistically stationary reference and perturbed
fields is directly balanced by the decorrelating action of the
internal production of uncertainty.

The uncertainty dissipation rate
$\left\langle\varepsilon_{\Delta}\right\rangle$ reaches its long-time
asymptotic balance with
$\left\langle\varepsilon_{\text{tot}}\right\rangle$,
i.e. $\left\langle\varepsilon_{\Delta}\right\rangle/\left\langle\varepsilon_{\text{tot}}\right\rangle>0.95$,
at about $\tau = 16.1$ for F3 and at about $\tau =
  5.6$ for both F1 and F2. This is slightly before but close to the
time $\tau = 6.5$ when $\left\langle
F_{\Delta}\right\rangle/\left\langle\varepsilon_{\Delta}\right\rangle=0.06$
stops being negligible in F1 and the perturbation evolutions start
diverging between F1 and F2. The presence of positive $\left\langle
F_{\Delta}\right\rangle$ in F1 delays the decay towards $0$ of ${\rm
  d}\left\langle E_{\Delta}\right\rangle/{\rm d}t$ which is reached at
about $\tau = 10.6$ for F1 but
$\tau = 8.9$ for $F2$. In the case of $F1$ one
might even argue that an approximate steady state has resulted for
${\rm d}\left\langle E_{\Delta}\right\rangle/{\rm d}t$ between
$\tau =6.5$ and $\tau=
  10.6$, the time range corresponding to the linear growth regime
perhaps observed in figure \ref{fig:time evolution of uncertinty
  case1} for F1 and also in some previous DNS
\citep{berera2018chaotic,boffetta2017chaos}. After
$\tau= 10.6$, $\left\langle
P_{\Delta}\right\rangle$ oscillates around zero, corresponding to the
saturation of $\left\langle E_{\Delta}\right\rangle/\left\langle
E_{\text{tot}}\right\rangle$ at a value
$0.97\pm0.07$ in figure \ref{fig:time evolution
  of uncertinty case1}. This reflects the total decorrelation between
the F1 reference and perturbed fields and leads to a long-time
saturation balance $\left\langle F_{\Delta}\right\rangle \approx
\left\langle \varepsilon_{\Delta} \right\rangle$ in F1 which is to be
contrasted with $\left\langle P_{\Delta}\right\rangle \approx
\left\langle \varepsilon_{\Delta} \right\rangle$ in F2 and F3. Note
that the long-time saturation is such that $\left\langle
F_{\text{corr}}\right\rangle\approx0$ and $\left\langle
F_{\Delta}\right\rangle\approx\left\langle
F^{(1)}+F^{(2)}\right\rangle$ in all cases, including F1. Hence, the
long-time saturation balance between $\left\langle
F_{\Delta}\right\rangle$ and $\left\langle
\varepsilon_{\Delta}\right\rangle$ in case F1 simply reflects
$\left\langle\varepsilon_{\text{tot}}\right\rangle\approx\left\langle
F^{(1)}+F^{(2)}\right\rangle$.

\begin{figure}
	\centering 
	\subfigure[case F1]{
		\label{fig:early spectra case1}
		\includegraphics[width=0.49\textwidth]{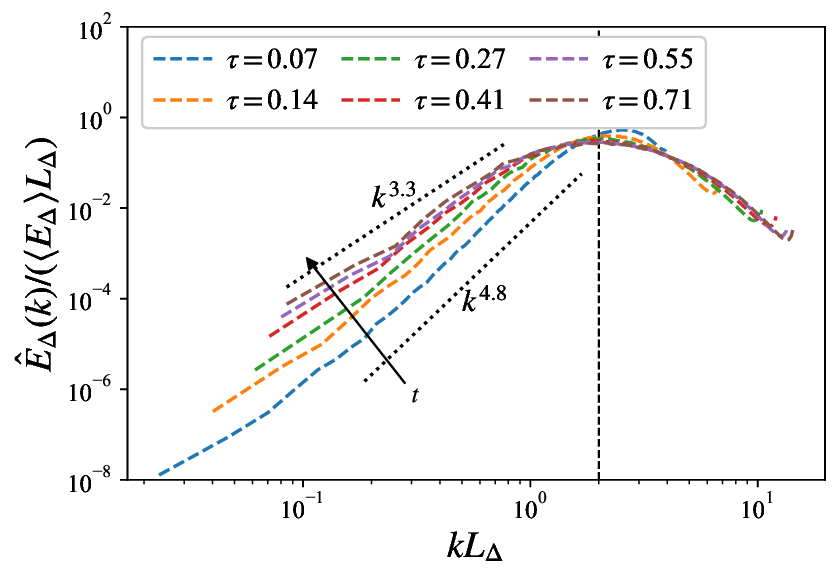}}
	\subfigure[case F2]{
		\label{fig:early spectra case2}
		\includegraphics[width=0.49\textwidth]{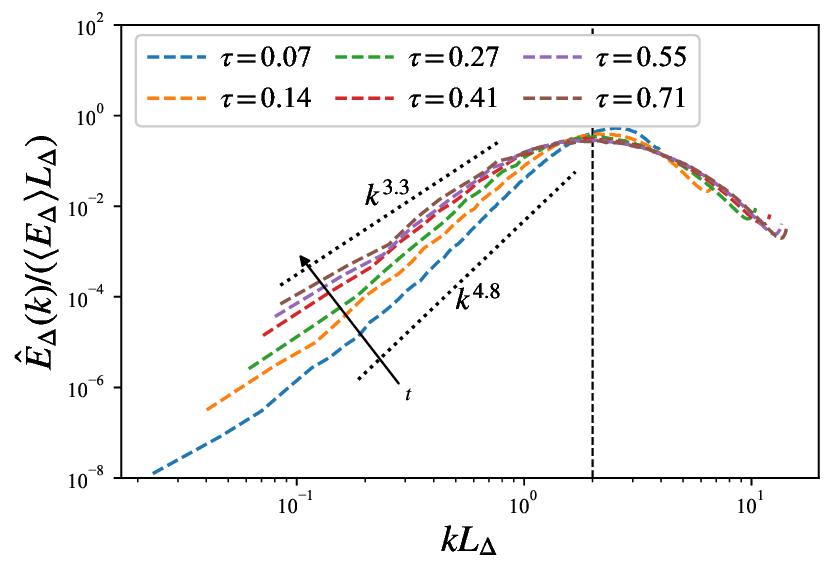}}
	\subfigure[case F3]{
		\label{fig:early spectra case3}
		\includegraphics[width=0.49\textwidth]{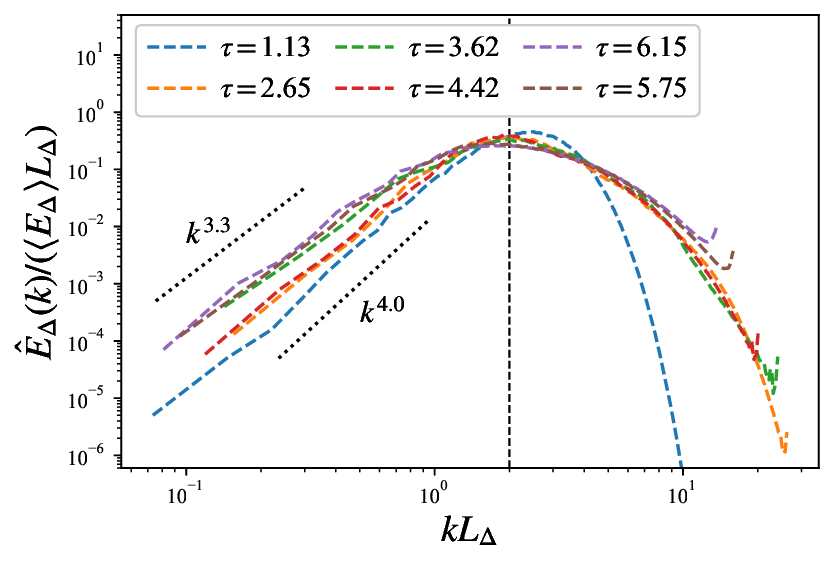}}
	\caption{Early-time evolution of uncertainty energy spectrum for different cases. The spectra are normalized by $\left\langle E_{\Delta}\right\rangle$ and $L_{\Delta}$.} 
	\label{fig:early spectra} 
\end{figure}

In figure \ref{fig:P-Epsilon} we concentrate on the time-evolution of
the production-dissipation ratio $\left\langle
P_{\Delta}\right\rangle/\left\langle\varepsilon_{\Delta}\right\rangle$
in all three cases F1, F2 and F3. As highlighted in the insets of this
figure's plots, there is, in all three cases, a time range when
$\left\langle
P_{\Delta}\right\rangle/\left\langle\varepsilon_{\Delta}\right\rangle$
is about constant, i.e. a time range when the evolutions of
$\left\langle P_{\Delta}\right\rangle$ and
$\left\langle\varepsilon_{\Delta}\right\rangle$ are similar. In all
three cases this time range includes the time range of exponential
growth of $\left\langle E_{\Delta}\right\rangle$ identified in the
previous subsection; in fact, in case F3 it more or less exactly
coincides with it. To be specific, $\left\langle
  P_{\Delta}\right\rangle/\left\langle\varepsilon_{\Delta}\right\rangle=1.61\pm0.03$
from $\tau = 0.6$ to $\tau =
  2.5$ for F1 and F2, and $\left\langle
P_{\Delta}\right\rangle/\left\langle\varepsilon_{\Delta}\right\rangle=1.61\pm0.08$
from $\tau = 7.2$ to $\tau = 12.8$ for F3. These two values are very
close (and the additional case F4 in Appendix
  \ref{app:Reynolds-number dependence of the time range of the
    exponential regime} returns a similar value for $\left\langle
  P_{\Delta}\right\rangle/\left\langle\varepsilon_{\Delta}\right\rangle$
  in F4's similarity regime), indicating that the similarity value of
the production-dissipation ratio $\left\langle
P_{\Delta}\right\rangle/\left\langle\varepsilon_{\Delta}\right\rangle$
might be universal and independent of Reynolds number, as the presence
of a strange attactor might perhaps imply.

	\begin{figure}
	\centering 
	\subfigure[case F1]{
		\label{fig:spectra case1}
		\includegraphics[width=0.49\textwidth]{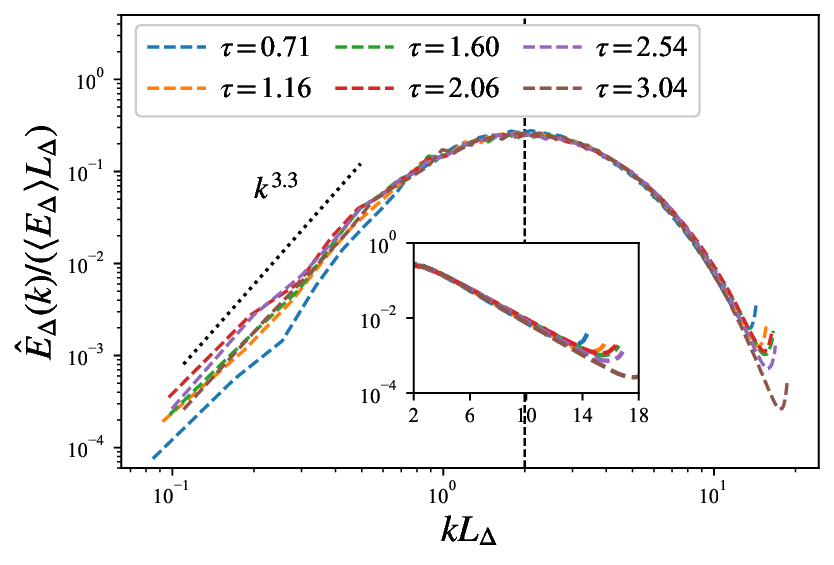}}
	\subfigure[case F2]{
		\label{fig:spectra case2}
		\includegraphics[width=0.49\textwidth]{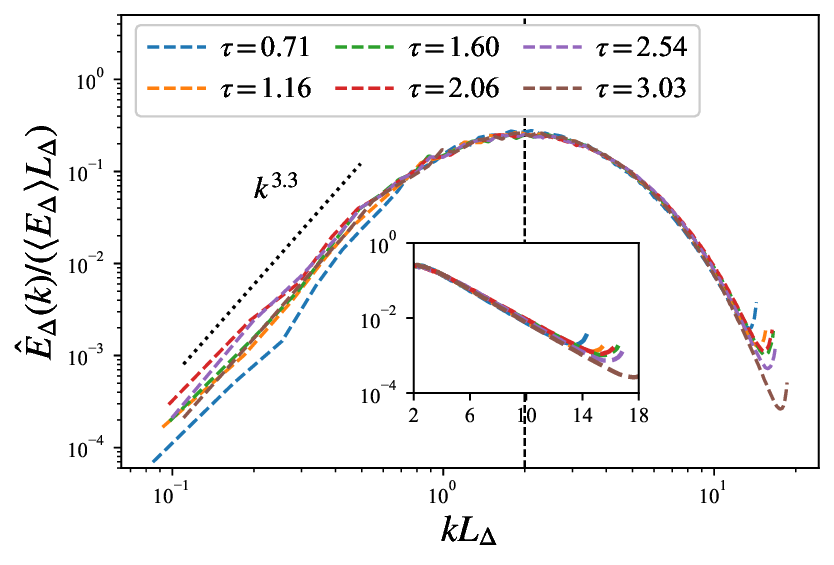}}
	\subfigure[case F3]{
		\label{fig:spectra case3}
		\includegraphics[width=0.49\textwidth]{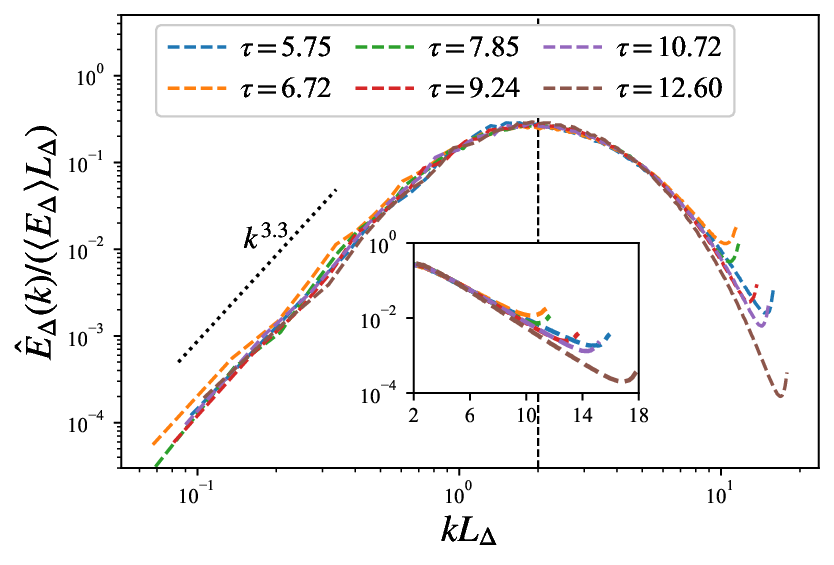}}
		\subfigure[comparison F1 - F2 - F3]{
		\label{fig:spectra comparsion}
		\includegraphics[width=0.49\textwidth]{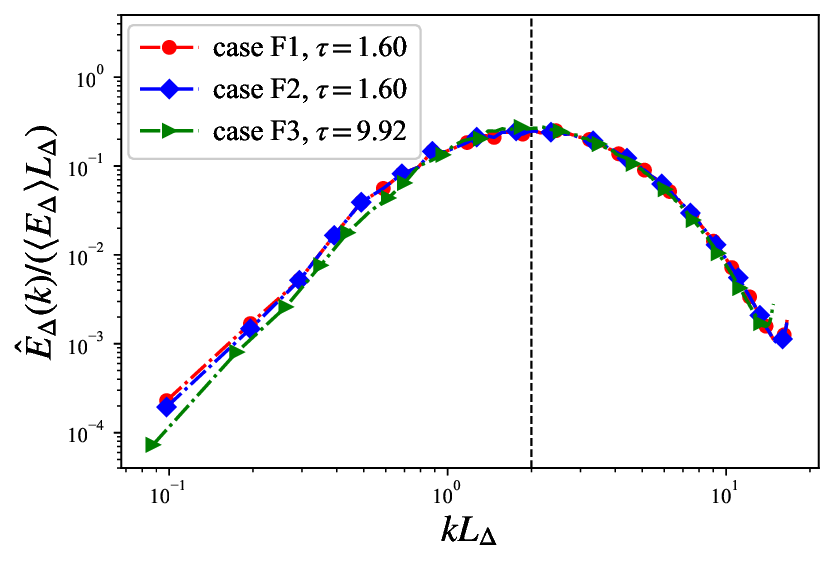}}
	\caption{Uncertainty energy spectra for cases (\textit{a}) F1,
          (\textit{b}) F2 and (\textit{c}) F3 in the similarity
          regime. The spectra are normalized by $\left\langle
          E_{\Delta}\right\rangle$ and $L_{\Delta}$. Inset:
          semilogarithmic plot of the uncertainty spectra in the
          wavenumber range higher than
          $2/L_{\Delta}$. The collapse of the
            normalized uncertainty spectra for cases F1, F2 and F3 in
            the similarity regime is shown in (\textit{d}).}
	\label{fig:spectra} 
\end{figure}

\subsubsection{Uncertainty spectrum}
The uncertainty dissipation rate is the integral over all wavenumbers
$k$ of $k^{2} \hat{E}_{\Delta}(k)$ where $\hat{E}_{\Delta}(k)$ is the
uncertainty spectrum, i.e. the energy spectrum of the velocity
difference field. The similarity in the evolutions of uncertainty
production and dissipation rates raises the question whether the
uncertainty spectrum evolves in some self-similar manner over the time
range of that similarity. We answer this question in terms of the
integral length scale of the velocity-difference fields considered
here which is $L_{\Delta}=\left(3\pi/4\left\langle
E_{\Delta}\right\rangle\right)\int k^{-1}\hat{E}_{\Delta}(k){\rm d}k$
(see \citet{batchelor1953theory} for an introduction to this length
scale for any statistically homogeneous/periodic velocity
field). $L_{\Delta}$ is a measure of the length over which the
velocity difference field is correlated, i.e. a characteristic length
scale of uncertainty containing eddies.

Soon after the initial decay of $\left\langle E_{\Delta}
\right\rangle$, the uncertainty spectra collapse with $\left\langle
E_{\Delta} \right\rangle (t)$ and $L_{\Delta} (t)$ at wavenumbers
larger than $2/L_{\Delta}$ as shown in figure \ref{fig:early spectra},
i.e. $\hat{E}_{\Delta}(k,t) = \left\langle E_{\Delta} \right\rangle
L_{\Delta} f(kL_{\Delta})$ for $kL_{\Delta}\ge 2$, where $f$ is a
dimensionless function of dimensionless wavenumber. At wavenumbers
$kL_{\Delta} < 1$ the energy spectra have an approximately power law
dependence on $k$ but do not collapse till soon after the time when
the exponential growth of $\left\langle E_{\Delta} \right\rangle (t)$
and the uncertainty's production-dissipation similarity ($\left\langle
P_{\Delta}\right\rangle/\left\langle\varepsilon_{\Delta}\right\rangle\approx
1.6-1.7$) sets in. Over the time range when $\left\langle
P_{\Delta}\right\rangle/\left\langle\varepsilon_{\Delta}\right\rangle\approx
1.6-1.7$, the uncertainty spectrum is self-similar, i.e. evolves as
$\hat{E}_{\Delta}(k,t) = \left\langle E_{\Delta} \right\rangle
L_{\Delta} f(kL_{\Delta})$ for all wavenumbers (see figure
\ref{fig:spectra}). The peak of the spectrum is at $k\approx
2/L_{\Delta}$ in all three cases F1, F2 and F3. At wavenumbers below
$2/L_{\Delta}$ the uncertainty spectra have an approximately $k^{3.3}$
power law shape, while at wavenumbers above $2/L_{\Delta}$, they
appear to have an exponential shape. Similar uncertainty spectrum
shapes have been found in a previous DNS study
\citep{berera2018chaotic}.

It is remarkable that the uncertainty spectrum is self-similar in case
F3 in the exact same way that it is self-similar in cases F1 and F2
over the time range where $\left\langle
P_{\Delta}\right\rangle/\left\langle\varepsilon_{\Delta}\right\rangle$
is approximately constant. In fact, the
  self-similar uncertainty spectrum even seems to be the same for F3,
  F1 and F2 as seen by the collapse in figure \ref{fig:spectra
    comparsion}, suggesting a universal shape for the self-similar
  uncertainty spectrum in HIT.  This is remarkable not only because
$F3$ has a very different Reynolds number and forcing than F1 and F2,
but more importantly because the F3 reference field is not
statistically stationary in that time range whereas the F1 and F2
reference fields are. In the F3 case, the uncertainty spectrum reaches
its self-similar state at $\tau\approx5.8$ and the reference field
becomes statistically stationary at $\tau=9.4$.

\begin{figure}
	\centering 
	\subfigure[case F1]{
		\label{fig:uncertainty spectra case1}
		\includegraphics[width=0.49\textwidth]{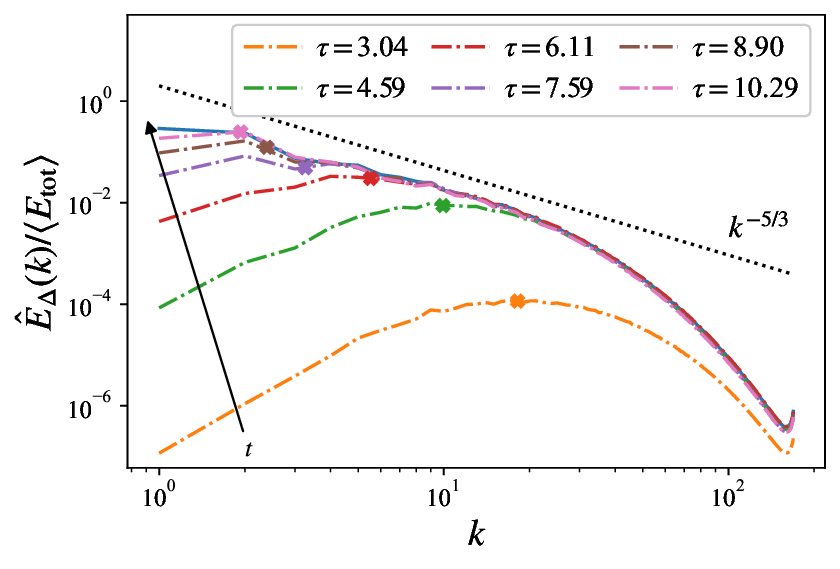}}
	\subfigure[case F2]{
		\label{fig:uncertainty spectra case2}
		\includegraphics[width=0.49\textwidth]{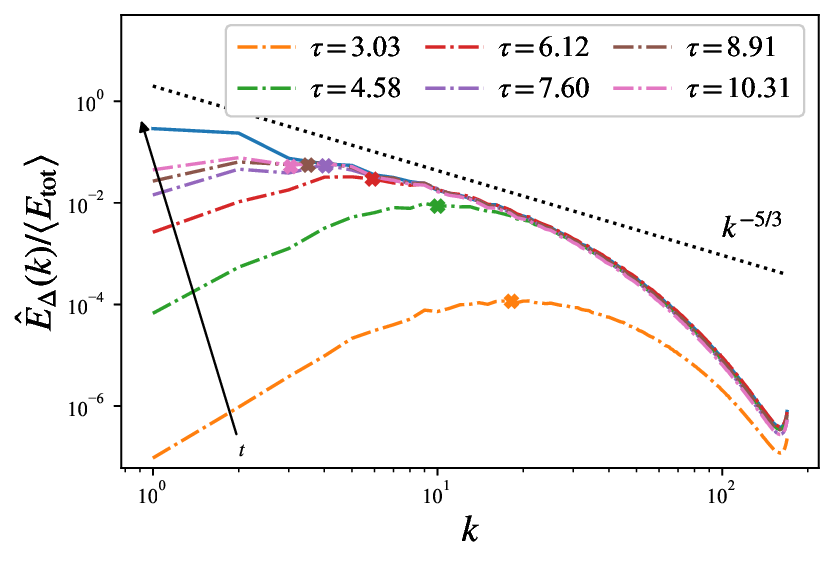}}
	\subfigure[case F3]{
		\label{fig:uncertainty spectra case3}
		\includegraphics[width=0.49\textwidth]{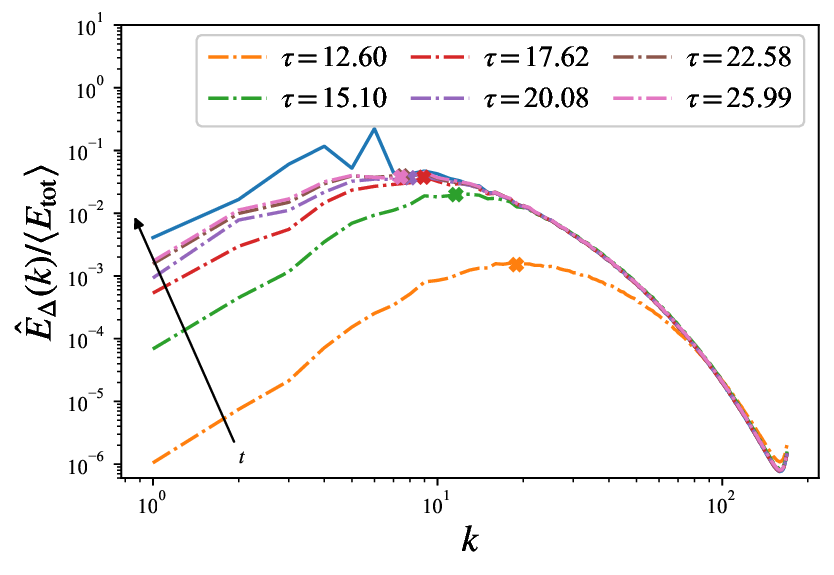}}
	\caption{Uncertainty spectra (dash lines) for different cases after the times when the reference and perturbed fields decorrelate at the largest resolvable wavenumber. The uncertainty spectra are normalized by $\left\langle E_{\text{tot}}\right\rangle$. The dots on the uncertainty spectra represent $k=2/L_{\Delta}$. The solid line represents the energy spectrum of the reference field when it is statistically steady. The energy spectrum is normalized by $\left\langle E^{(1)}\right\rangle$. The uncertainty spectrum shifts gradually along with the arrows representing the direction of time advance. } 
	\label{fig:uncertainty spectra} 
\end{figure}

After the time-range where $\left\langle
P_{\Delta}\right\rangle/\left\langle\varepsilon_{\Delta}\right\rangle$
is approximately constant, the uncertainty spectrum is no longer
self-similar (see figure \ref{fig:uncertainty spectra}). This happens
at $\tau=3.5$ for cases F1 and F2 and at
$\tau=12.6$ for case F3 when
$\hat{E}_{\Delta}(k_{max})/\hat{E}_{\text{tot}}(k_{max})>0.95$. These
are the times when the reference and perturbed fields decorrelate at
the largest resolvable wavenumber (see figure \ref{fig:uncertainty
  spectra}). The process of decorrelation between the two fields
proceeds by decorrelating them at progressively smaller wavenumbers,
causing the uncertainty spectrum to collapse with the reference
field's energy spectrum over a progressively wider range of the higher
wavenumbers (see figure \ref{fig:uncertainty spectra}).  This
progressive decorrelation process from high to small wavenumbers and
the uncertainty spectrum's progressive convergence towards the
reference field's spectrum prevents the uncertainty spectrum from
being self-similar. For F1, the uncertainty spectrum finally collapses
with the reference field's energy spectrum at all wavenumbers,
indicating that the two fields eventually decorrelate completely at
all wavenumbers (see figure \ref{fig:uncertainty spectra case1}). The
same happens for F2 and F3 except over the wavenumbers acted by the
forcing where a gap always remains between the uncertainty and the
reference field spectra, indicating that the two fields retain a
degree of correlation at these large scales (see figure
\ref{fig:uncertainty spectra case2}, \ref{fig:uncertainty spectra
  case3}).

\begin{figure}
	\centering 
	\subfigure[case F1]{
		\label{fig:LDelta case1}
		\includegraphics[width=0.49\textwidth]{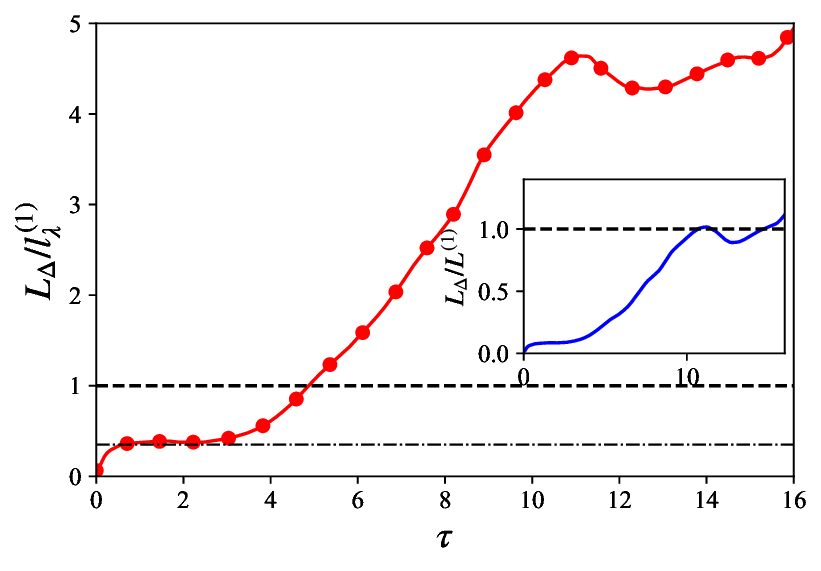}}
	\subfigure[case F2]{
		\label{fig:LDelta case2}
		\includegraphics[width=0.49\textwidth]{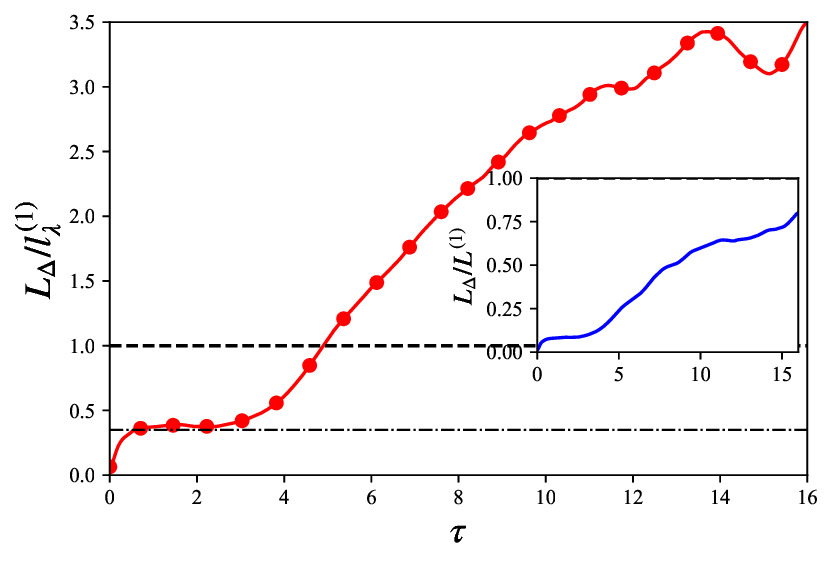}}
	\subfigure[case F3]{
		\label{fig:LDelta case3}
		\includegraphics[width=0.49\textwidth]{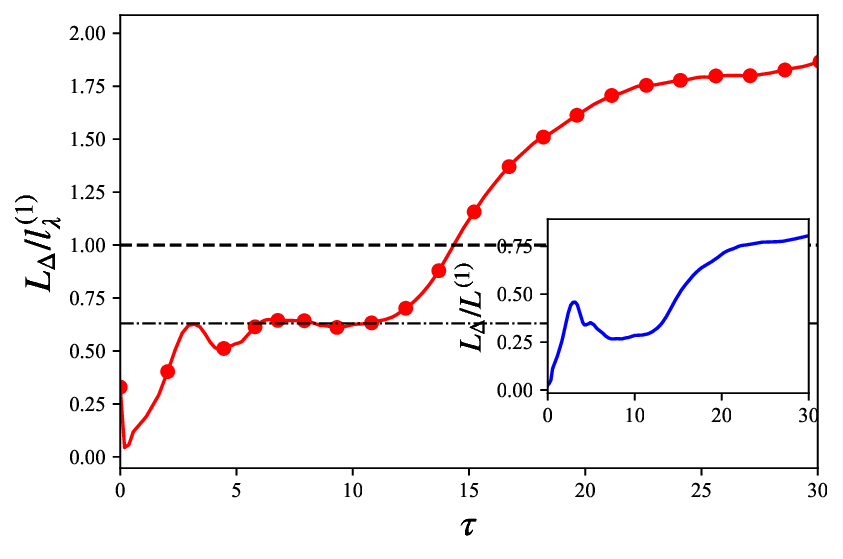}}
	\subfigure[comparison F1 - F2]{
		\label{fig:LDelta compare}
		\includegraphics[width=0.49\textwidth]{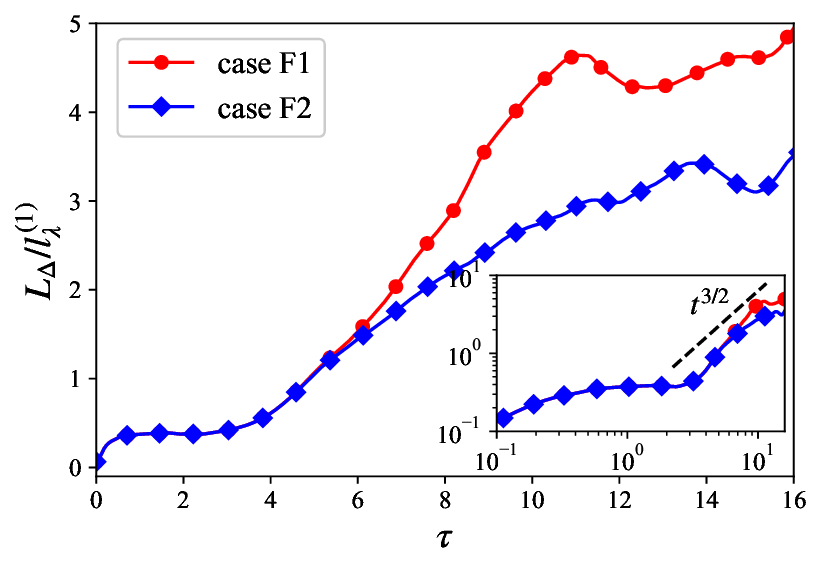}}
	\caption{Time evolution of $L_{\Delta}/l_{\lambda}^{(1)}$ for different cases. Inset: time evolution of
		$L_{\Delta}/L^{(1)}$. The time evolutions of $L_{\Delta}/l_{\lambda}^{(1)}$ in cases F1 and F2 are plotted together in (\textit{d}). Inset: the log-log plot of the time evolutions of $L_{\Delta}/l_{\lambda}^{(1)}$.}
	\label{fig:LDelta} 
\end{figure}

\subsubsection{Characteristic length of uncertainty}

The growth of $L_{\Delta}$ is evident in figure \ref{fig:uncertainty
  spectra}. We therefore plot its time evolution in figure
\ref{fig:LDelta} and compare it with the integral and Taylor length
scales ($L$ and $l_{\lambda}$ respectively) of the reference field for
each case F1, F2 and F3. At the very early times when uncertainty
dissipation dominates, the velocity-difference field decays and its
integral length scale normalised by $L$ is, correspondingly,
increasing. In the stationary turbulence F1 and F2 cases, this time
regime is followed by the chaotic regime where $\left\langle
E_{\Delta}\right\rangle$ grows exponentially and where
$L_{\Delta}/l_{\lambda}^{(1)}$ remains relatively constant at
$0.38\pm0.01$. A constant
$L_{\Delta}/l_{\lambda}^{(1)}$ (though a different constant,
$L_{\Delta}/l_{\lambda}^{(1)} = 0.64\pm0.03$) is also observed in the
non-stationary F3 case during the chaotic regime even though
$l_{\lambda}^{(1)}$ grows in time for some of that regime and even
though this time regime does not follow immediately after the
dissipation-dominated regime. In fact, $L_{\Delta}$ decreases between
the dissipation-dominated and the chaotic regime in the F3 case. It is
noteworthy that $L_{\Delta}$ reaches $l_{\lambda}^{(1)}$ at
$\tau=4.9$ for cases F1 and F2 and at $\tau=14.5$
for case F3, a little before the average uncertainty dissipation rate
reaches its stationary value in figure \ref{fig:uncertainty equation},
i.e. $\tau \approx 5.6$ for F1 and F2 and $\tau
\approx 16.1$ for F3. The link between $L_{\Delta}$ and the Taylor
length of the reference field is potentially interesting as the Taylor
length is the mean distance between stagnation points in a homogeneous
isotropic turbulence \citep{goto2009dissipation} and therefore tends
to represent the average size of turbulent eddies which is highly
weighted towards the more numerous smallest ones.

Following the exponential growth of $\left\langle E_{\Delta}\right\rangle$, three consecutive time regimes follow for F1 and F2. First, one observes an approximately power-law growth of $L_{\Delta}$, identical for both F1 and F2 as shown in figure \ref{fig:LDelta compare}, more or less coinciding with the exponential of exponential growth of $\left\langle E_{\Delta}\right\rangle$ till $\tau=6.5$. In this time regime, $L_{\Delta}\sim t^{3/2}$ is a good fit. This fit is reminiscent of the power-law growth of the predictability scale $k_{E}^{-1}\sim t^{3/2}$ obtained in previous numerical simulations \citep{boffetta2017chaos,leith1972predictability} and theoretical arguments \citep{boffetta2017chaos,lorenz1969predictability,frisch1995turbulence}, as a companion conclusion to the linear growth of $\left\langle E_{\Delta}\right\rangle$. However, $L_{\Delta}$ and $k_{E}^{-1}$ are not equivalent: the predictability scale is defined as the inverse of the minimum wavenumber $k_E$ such that $\hat{E_{\Delta}}(k_{E})/\hat{E}_{\text{tot}}(k_{E})=1$, and $k_{E}^{-1}\sim t^{3/2}$ is obtained on the assumption that the decorrelation process happens in the inertial range. $L_{\Delta}\sim t^{3/2}$ is observed without concurrent linear growth of $\left\langle E_{\Delta}\right\rangle$ but a concurrent exponential of exponential $\left\langle E_{\Delta}\right\rangle$ growth instead.

The second consecutive regime which follows for F1 and F2 is an
apparently linear growth of $L_{\Delta}$ that lasts till the time when
$L_{\Delta}$ saturates to a constant. The third and final regime is
this approximately constant $L_{\Delta}$ regime where
$L_{\Delta}\approx L^{(1)}$ for F1 (see figure \ref{fig:LDelta case1})
in agreement with the eventual complete decorrelation of the reference
and perturbed fields and where $L_{\Delta}\approx
  (0.70\pm0.06)L^{(1)}$ (smaller than $L^{(1)}$) for F2 (see figure
\ref{fig:LDelta case2}) in agreement with the eventual partial
correlation between these two fields in this case.

In the F3 case, the chaotic regime where $\left\langle E_{\Delta}\right\rangle$ grows exponentially and $L_{\Delta}/l_{\lambda}^{(1)} = 0.64\pm0.03$ is followed by an intermediate regime where $L_{\Delta}$ grows to eventually reach the final constant regime where $L_{\Delta} = (0.81\pm0.03) L^{(1)}$ (smaller than $L^{(1)}$) characterising the final saturation (see figure \ref{fig:LDelta case3}). As for F2, the fact that $L_{\Delta}$ is significantly lower than $L$ in the eventual saturation regime reflects the partial long-time correlation imposed by the identical forcing in the reference and perturbed fields.

\subsection{\label{sec:Quantitative analysis of the uncertainty growth}Quantitative analysis of the uncertainty growth}
\subsubsection{\label{sec:From similarity to exponential growth}From similarity to exponential growth}
\begin{figure}
	\centering \subfigure[case F1]{
		\label{fig:approximation production case1}
		\includegraphics[width=0.49\textwidth]{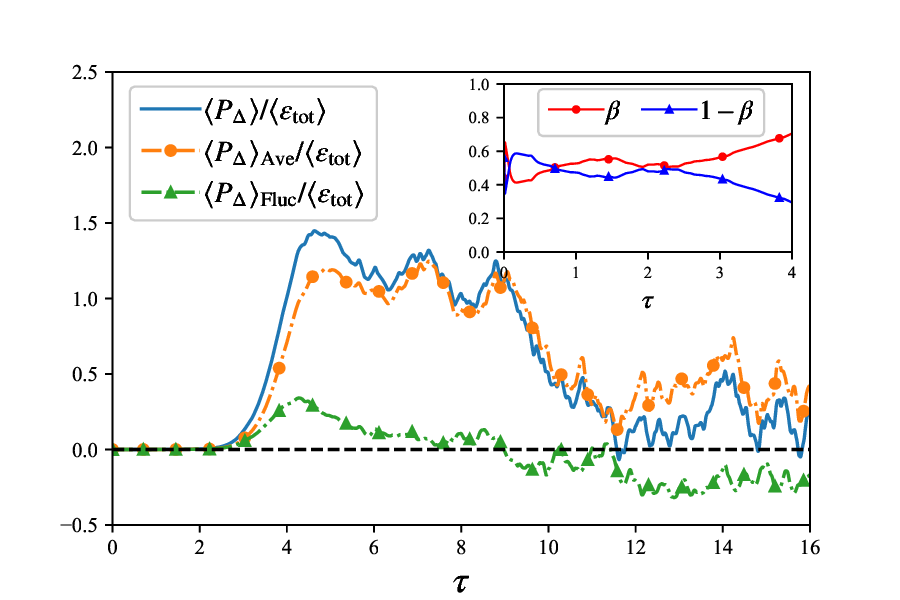}}
	\subfigure[case F2]{
		\label{fig:approximation production case2}
		\includegraphics[width=0.49\textwidth]{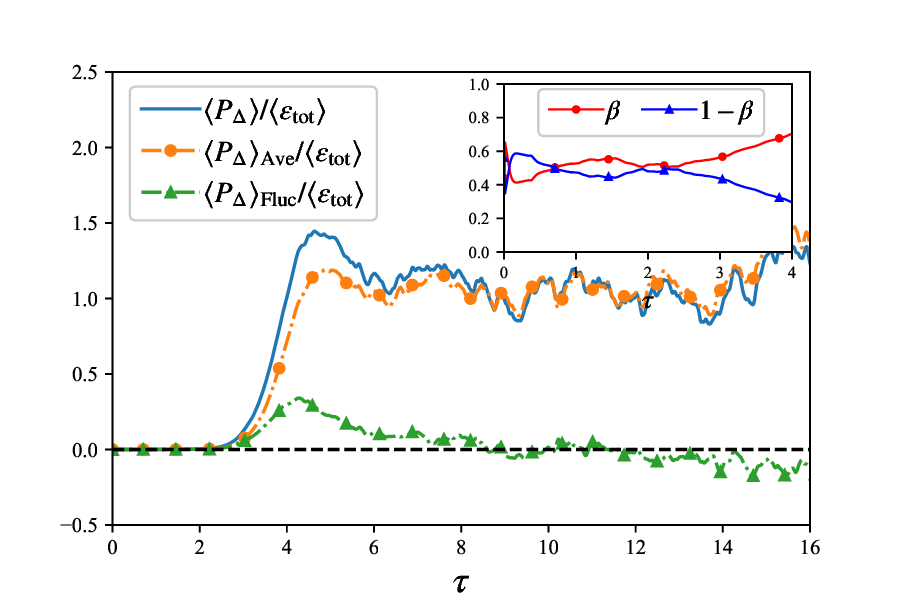}}
	\subfigure[case F3]{
		\label{fig:approximation production case3}
		\includegraphics[width=0.49 \textwidth]{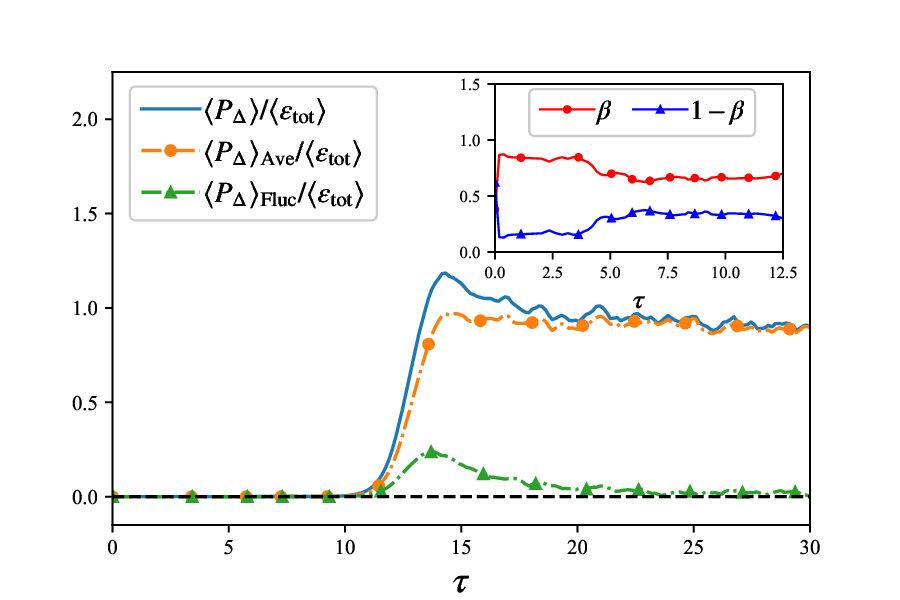}}
	\caption{Time evolution of $\left\langle P_{\Delta}\right\rangle$ and its decomposition into average and fluctuation parts for different cases. Inset: the early-time evolution of the ratio of average term/total production and fluctuation term/total production.} 
	\label{fig:approximation production} 
\end{figure}

\begin{figure}
	\centering 
	\subfigure[case F1]{
		\label{fig:eigenvalue case1}
		\includegraphics[width=0.49\textwidth]{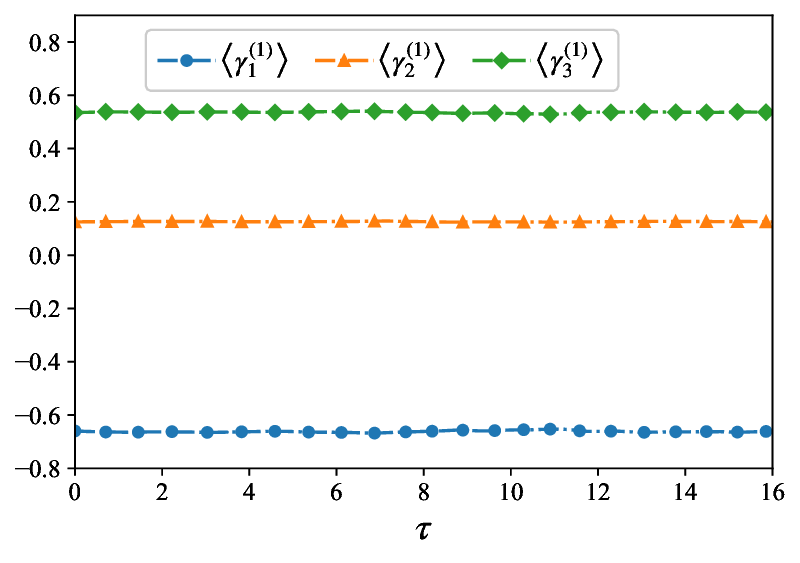}}
	\subfigure[case F2]{
		\label{fig:eigenvalue case2}
		\includegraphics[width=0.49\textwidth]{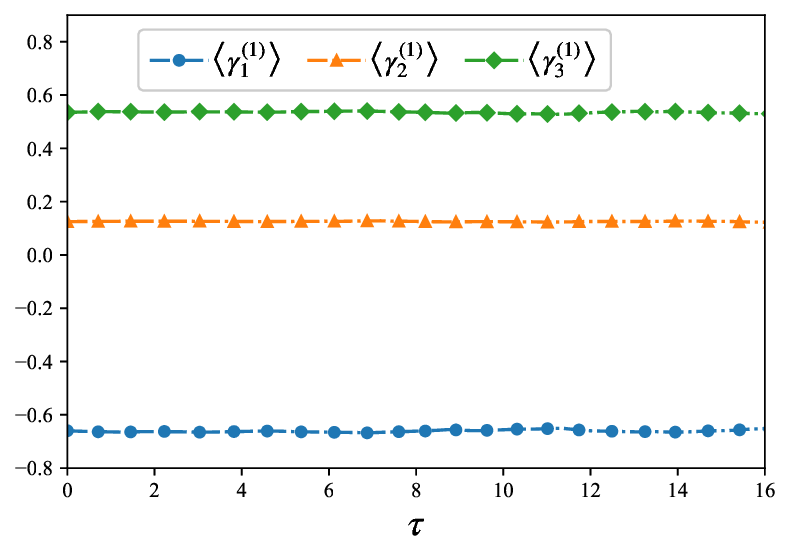}}
	\subfigure[case F3]{
		\label{fig:eigenvalue case3}
		\includegraphics[width=0.49\textwidth]{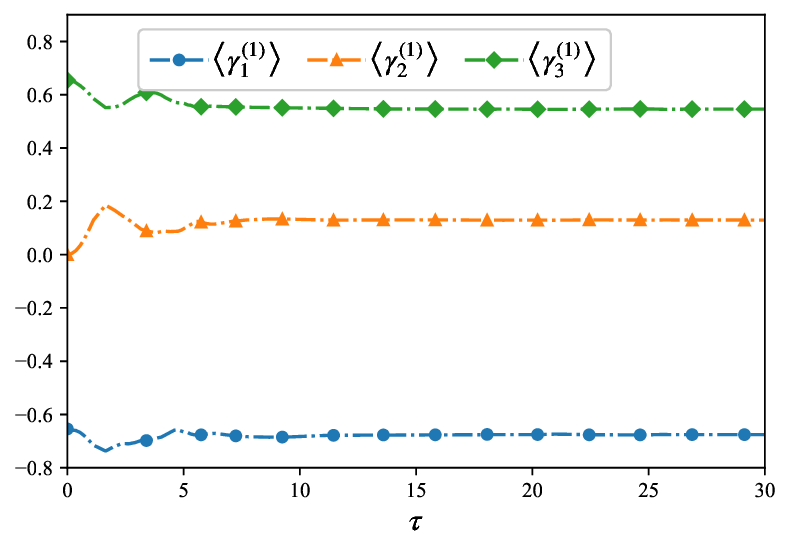}}
	\caption{Time evolution of $\left\langle\gamma_{i}\right\rangle$ in the reference flows for different cases.} 
	\label{fig:eigenvalue} 
\end{figure}

\begin{figure}
	\centering 
	\subfigure[case F1]{
		\label{fig:Anistropy case1}
		\includegraphics[width=0.49\textwidth]{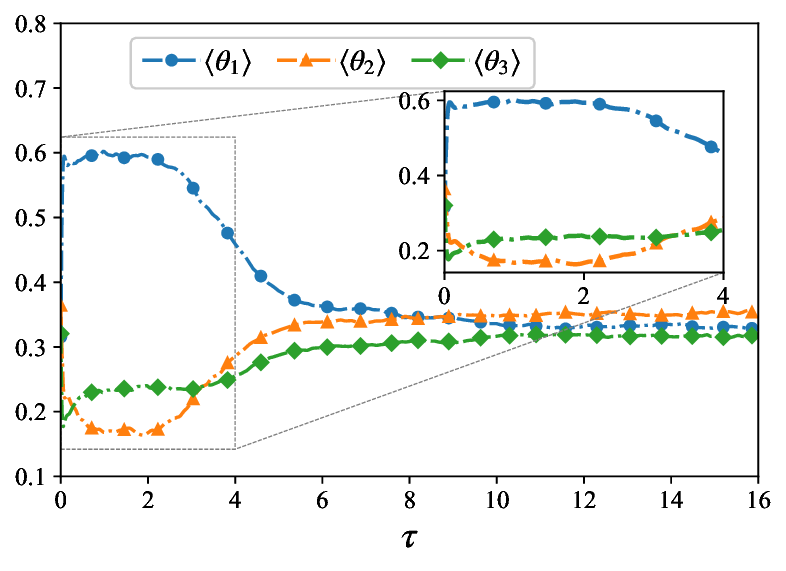}}
	\subfigure[case F2]{
		\label{fig:Anistropy case2}
		\includegraphics[width=0.49\textwidth]{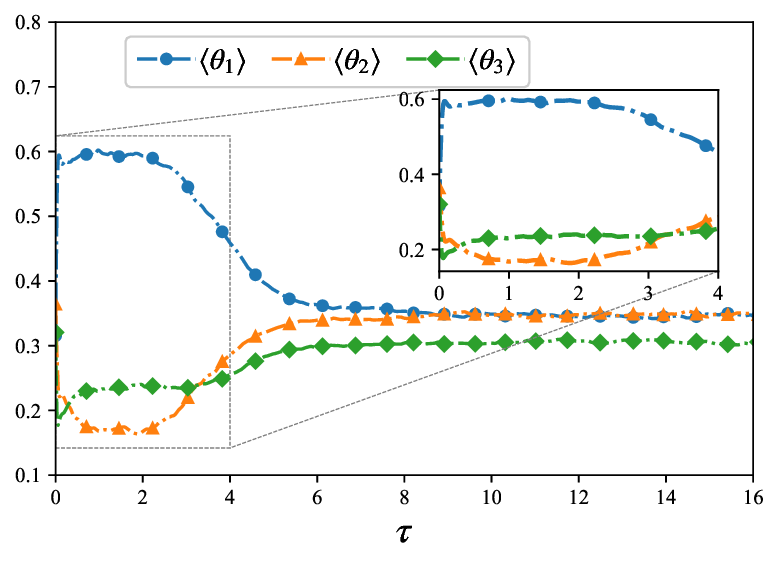}}
	\subfigure[case F3]{
		\label{fig:Anistropy case3}
		\includegraphics[width=0.49\textwidth]{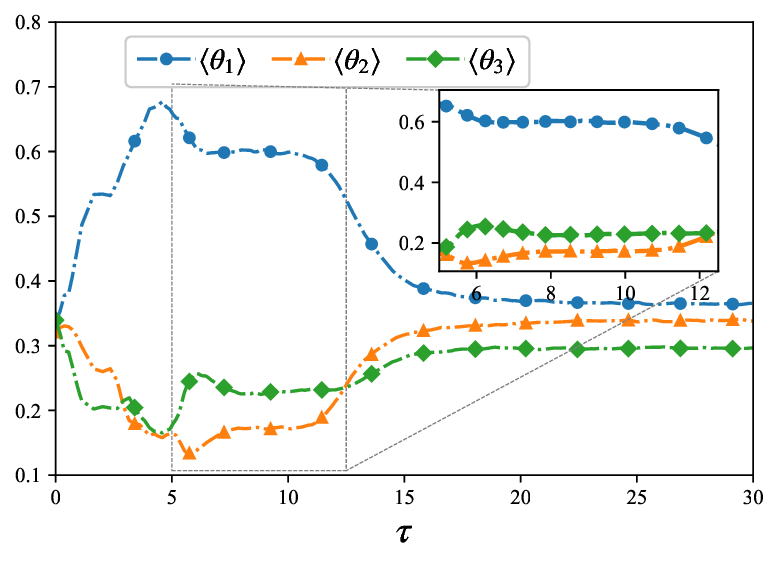}}
	\caption{Time evolution of $\left\langle\theta_{i}\right\rangle$ for different cases. Inset: the time evolution of $\left\langle\theta_{i}\right\rangle$ during the similarity regime.} 
	\label{fig:Anistropy} 
\end{figure}

\begin{figure}
	\centering
        \includegraphics[width=0.6\textwidth]{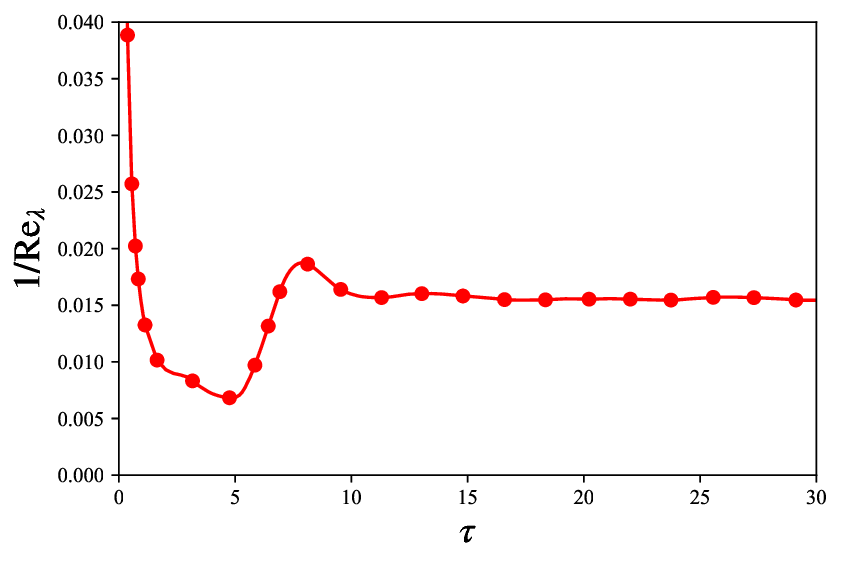}
	\caption{Time evolution of $1/\text{Re}_{\lambda}$ in case F3.} 
	\label{fig:reynoldsTaylorF3} 
\end{figure}

\begin{figure}
	\centering 
	\subfigure[case F1]{
		\label{fig:gamma case1}
		\includegraphics[width=0.49\textwidth]{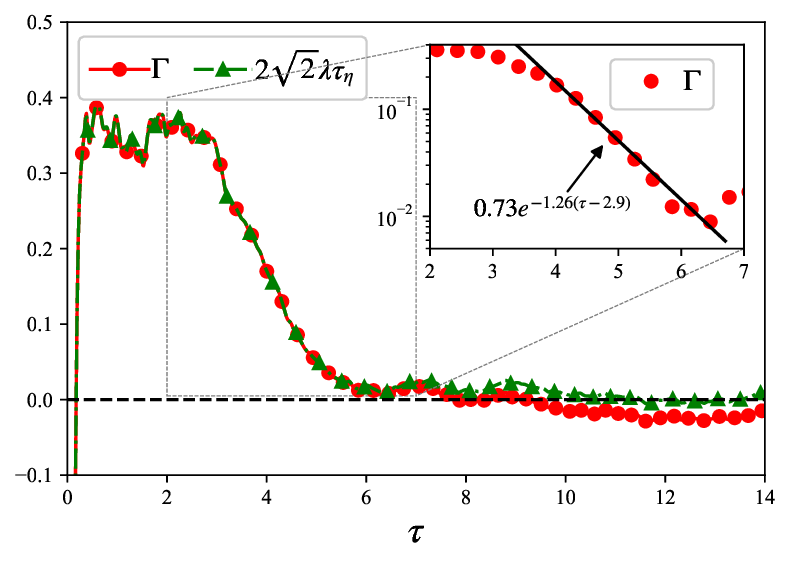}}
	\subfigure[case F2]{
		\label{fig:gamma case2}
		\includegraphics[width=0.49\textwidth]{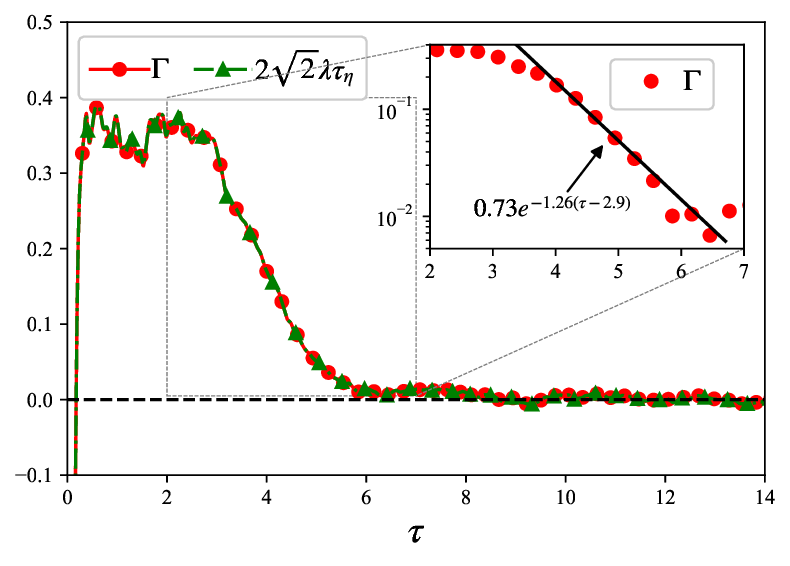}}
	\subfigure[case F3]{
		\label{fig:gamma case3}
		\includegraphics[width=0.49\textwidth]{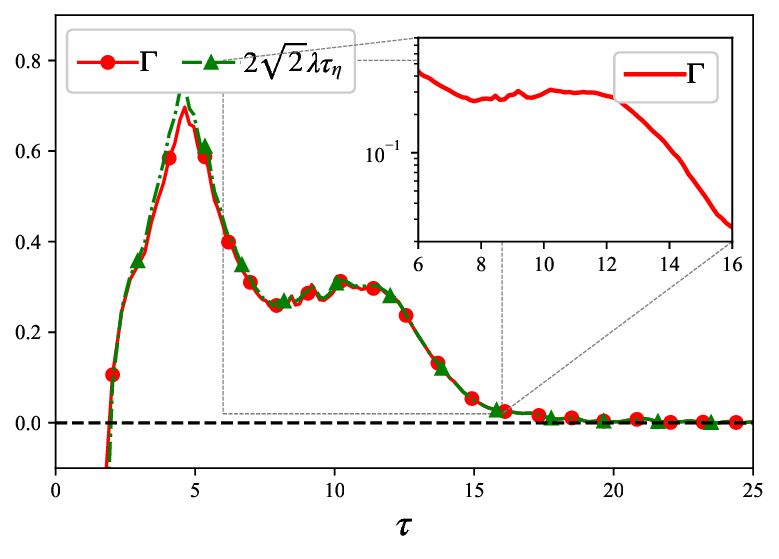}}
	\caption{Time evolution of $\Gamma$ and $2\sqrt{2}\lambda\tau_{\eta}$for different cases. Inset: the time evolution of $\Gamma$ during the similarity regime in semilogarithmic plot. The exponential function fit is indicated by a dash-dot line for F1 and F2.} 
	\label{fig:gamma} 
\end{figure}

\begin{figure}
	\centering 
	\includegraphics[width=0.65\textwidth]{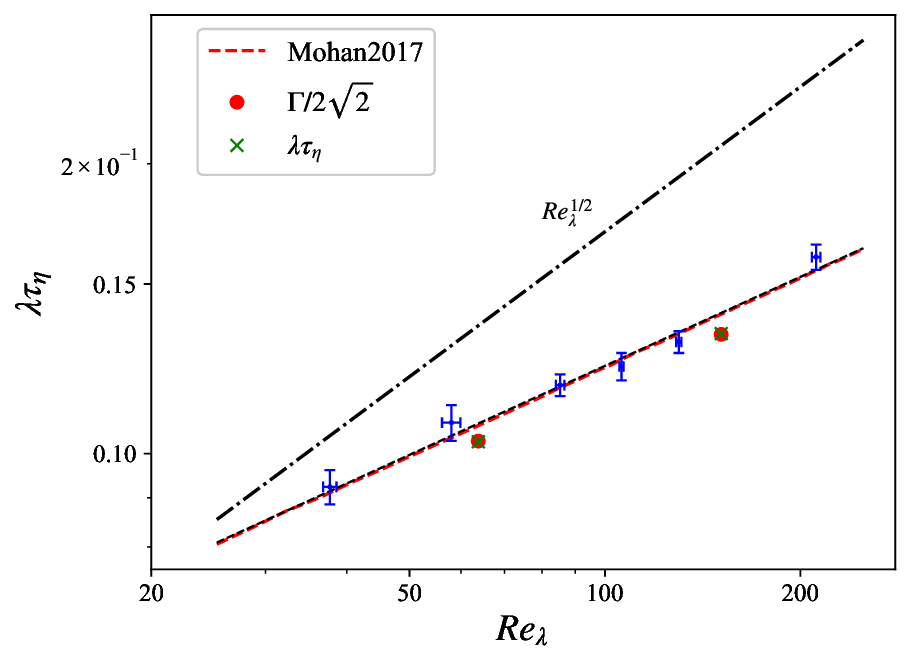}
	\caption{Schematic log-log plot of $\lambda\tau_{\eta}$ with the Taylor
		length-based Reynolds number $\text{Re}_{\lambda}$ according to the numerical results (in blue) and models calibrated with Bayesian inference (in red) of \citet{mohan2017scaling}. The Lyapunov exponents and the coefficient $\Gamma$ obtained in the present work (in green crosses and red points) are also plotted.} 
	\label{fig:Mohan} 
\end{figure}

When $\left\langle F_{\Delta} \right\rangle$ is identically zero (as in F2 and F3) or negligibly small compared to $\left\langle P_{\Delta}\right\rangle$ and $\left\langle\varepsilon_{\Delta}\right\rangle$ (as in F1 for $\tau$ smaller than about $6.5$) the evolution equation for $\left\langle E_{\Delta} \right\rangle$ becomes
\begin{equation}
\label{eq:first simplification}
\frac{{\rm d~}}{{\rm d} t}\left\langle E_{\Delta}\right\rangle=\left\langle P_{\Delta}\right\rangle-\left\langle \varepsilon_{\Delta}\right\rangle.
\end{equation}
To estimate $\left\langle P_{\Delta}\right\rangle$ in terms of $\left\langle E_{\Delta}\right\rangle$ and obtain an equation of the same form as equation (\ref{eq:exponential growth of uncertainty}), we apply a Reynolds decomposition to equation (\ref{eq:Production in principal axe}) and write
\begin{equation}
\label{eq:reynolds decomposition of production}
\left\langle
P_{\Delta}\right\rangle=-\sum_{i=1}^{3}\left\langle\Lambda^{(1)}_{i}\Delta
w_{i}^{2}\right\rangle=\underbrace{-\sum_{i=1}^{3}\left\langle\Lambda^{(1)}_{i}\right\rangle\left\langle\Delta
w_{i}^{2}\right\rangle}_{\left\langle
P_{\Delta}\right\rangle_{\text{Ave}}}\underbrace{-\sum_{i=1}^{3}\left\langle\Lambda_{i}^{(1)\prime}\Delta
{w_{i}^{2}}^{\prime}\right\rangle}_{\left\langle
P_{\Delta}\right\rangle_{\text{Fluc}}},
\end{equation}
where $\Lambda_{i}^{(1)\prime} \equiv
\Lambda^{(1)}_{i}-\left\langle\Lambda^{(1)}_{i}\right\rangle$ and
$\Delta {w_{i}^{2}}^{\prime} \equiv \Delta
w_{i}^{2}-\left\langle\Delta w_{i}^{2}\right\rangle$. In all cases F1,
F2 and F3, and at times after the similarity regime, the first term on
the right hand side of equation (\ref{eq:reynolds decomposition of
  production}) dominates over the second term and contributes the most
to $\left\langle P_{\Delta}\right\rangle$ (see figure
\ref{fig:approximation production}). During the part of the similarity
regime when $\left\langle E_{\Delta}\right\rangle$ grows
exponentially, $\beta \equiv \left\langle
P_{\Delta}\right\rangle_{\text{Ave}}/\left\langle
P_{\Delta}\right\rangle$ is constant in time and so is $1-\beta
=\left\langle P_{\Delta}\right\rangle_{\text{Fluc}}/\left\langle
P_{\Delta}\right\rangle$ (see insets of figure \ref{fig:approximation
  production}): $\beta$ is a constant equal to 
$0.53\pm0.02$ for F1 and F2 and equal to a slightly different value $0.66\pm0.02$ for
F3 where the Taylor length-based Reynolds number is significantly
lower than for F1 and F2. One may indeed expect the fluctuation
contribution $\left\langle P_{\Delta}\right\rangle_{\text{Fluc}}$ to
increase in magnitude with increasing Reynolds number relative to the
mean contribution $\left\langle P_{\Delta}\right\rangle_{\text{Ave}}$
in equation (\ref{eq:reynolds decomposition of production}), and
$\beta$ to therefore be a decreasing function of Reynolds number.

Defining $\gamma^{(1)}_{i} \equiv\Lambda^{(1)}_{i}/\sqrt{\left\langle\left|S^{(1)}_{ij}\right|^{2}\right\rangle}$ (where $\left|S_{ij}\right| \equiv \sqrt{S_{ij}S_{ij}}$) and $\theta_{i} \equiv \Delta w_{i}^{2}/2\left\langle
E_{\Delta}\right\rangle$, and using $\beta \equiv \left\langle
P_{\Delta}\right\rangle_{\text{Ave}}/\left\langle P_{\Delta}\right\rangle$, we have
\begin{equation}
\label{eq:approximation ODE of production}
\left\langle P_{\Delta}\right\rangle=-2\frac{\sum_{i=1}^{3}\left\langle\gamma^{(1)}_{i}\right\rangle\left\langle\theta_{i}\right\rangle}{\beta}\sqrt{\left\langle\left|S^{(1)}_{ij}\right|^{2}\right\rangle}\left\langle E_{\Delta}\right\rangle.
\end{equation}
We now examine the behaviours of $\left\langle\gamma^{(1)}_{i}\right\rangle$ and $\left\langle\theta_{i}\right\rangle$. 

We start with $\left\langle\gamma^{(1)}_{i}\right\rangle$ which,
unlike $\beta$ and $\left\langle\theta_{i}\right\rangle$, are
properties of the reference field and not of the velocity-difference
field: $\left\langle\gamma^{(1)}_{i}\right\rangle$ are the average
strain rates along the principal axes of the reference field's strain
rate tensor and they are plotted versus time in figure
\ref{fig:eigenvalue}. Note the constraints
$\sum_{i=1}^{3}\left\langle\gamma^{(1)}_{i}\right\rangle=0$ and
$\sum_{i=1}^{3}\left\langle\gamma_{i}^{(1)2}\right\rangle=1$. In cases
F1 and F2, where the reference flow is statistically stationary,
$\left\langle\gamma^{(1)}_{i}\right\rangle$ are constant in time and
$\left\langle\gamma^{(1)}_{1}\right\rangle:\left\langle\gamma^{(1)}_{2}\right\rangle:\left\langle\gamma^{(1)}_{3}\right\rangle\approx-0.65:0.12:0.53$
in agreement with \citet{betchov1956inequality}'s theoretical
demonstration that there must be one principal
  axis direction which is compressive on average and two which are on
  average stretching. In case F3, the reference flow is not
statistically stationary till about $\tau = 9.4$ but
$\left\langle\gamma^{(1)}_{i}\right\rangle$ acquire a stable value
before that and are already constant during the similarity period
$\tau \approx 5.8$ to $\tau \approx 12.60$ (see figure
\ref{fig:eigenvalue case3}). In case F3, we observe
$\left\langle\gamma^{(1)}_{1}\right\rangle:\left\langle\gamma^{(1)}_{2}\right\rangle:\left\langle\gamma^{(1)}_{3}\right\rangle\approx-0.68:0.13:0.55$,
which is very close to F1 and F2, also in
  agreement with \citet{betchov1956inequality}'s prediction.

The average uncertainty energy $\left\langle E_{\Delta}\right\rangle$
consists of three average uncertainty energies $\left\langle \Delta
w_{i}^{2}/2\right\rangle$ in the principal axes of the reference
field's strain rate tensor: $\left\langle E_{\Delta}\right\rangle =
\sum_{i=1}^{3} \left\langle \Delta w_{i}^{2}/2\right\rangle$. The
ratios $\left\langle\theta_{i}\right\rangle$ represent the proportion
of average uncertainty energy in each principal direction and they of
course sum up to 1. Their time evolution is shown in figure
\ref{fig:Anistropy}. Most of the uncertainty energy is concentrated in
the compressive direction till $\tau \approx8 -
  9$ in cases F1 and F2 and for all time in case F3, in agreement
with our observation at the end of subsection \ref{sec:Production
  term} that the production of uncertainty occurs by compressive
motions. At saturation times there is a tendency for equipartition of
average uncertainty energy in the three principal directions, in
particular for F1 where the reference and principal fields completely
decorrelate in the long term. The tendency remains for F2 and F3 but
the average uncertainty energy in the most stretching direction
remains significantly below the average uncertainty energy in the
other two directions thereby ensuring that $\left\langle
P_{\Delta}\right\rangle$ remains positive and the reference and
perturbation fields remain partially correlated during eventual
saturation.

In all three cases F1, F2 and F3,
$\left\langle\theta_{i}\right\rangle$ are approximately constant
during the similarity regime where
$\beta$ is also constant in time. During the similarity regime, the
$\left\langle\theta_{i}\right\rangle$ values are
$\left\langle\theta_{1}\right\rangle:\left\langle\theta_{2}\right\rangle:\left\langle\theta_{3}\right\rangle\approx0.58:0.19:0.23$
for cases F1 and F2 and
$\left\langle\theta_{1}\right\rangle:\left\langle\theta_{2}\right\rangle:\left\langle\theta_{3}\right\rangle\approx0.59:0.18:0.23$
for case F3. The values of $\left\langle\theta_{i}\right\rangle$
appear to be universal during the similarity regime whereas $\beta$
seems to be dependent on Reynolds number.

Finally we discuss the relation between
  $\left\langle\varepsilon_{\Delta}\right\rangle$ and $\left\langle
  P_{\Delta}\right\rangle$. The self-similar uncertainty spectrum
  $\hat{E}_{\Delta}(k,t) = \left\langle E_{\Delta} \right\rangle
  L_{\Delta} f(kL_{\Delta})$ implies that the uncertainty dissipation
  is
\begin{equation}
	\label{eq:uncertainty dissipation}
	\left\langle\varepsilon_{\Delta}\right\rangle=2\nu\int
        k^{2}\hat{E}_{\Delta}(k)\mathrm{d}k=2\nu\left\langle
        E_{\Delta} \right\rangle L_{\Delta}\int
        k^{2}f(kL_{\Delta})\mathrm{d}k=2\nu\frac{\left\langle
          E_{\Delta} \right\rangle}{L_{\Delta}^{2}}\int
        x^{2}f(x)\mathrm{d}x,
\end{equation}
where $\int x^{2}f(x)dx$ is a time-constant. Defining $\alpha \equiv
\left\langle\varepsilon_{\Delta}\right\rangle/\left\langle
P_{\Delta}\right\rangle$, we obtain, from equation
(\ref{eq:approximation ODE of production}) and (\ref{eq:uncertainty
  dissipation})
\begin{equation}
	\label{eq:alpha}
	\alpha=-\left[ \frac{\beta\int_{x}x^{2}f(x)\mathrm{d}x}{\sum_{i=1}^{3}\left\langle\gamma^{(1)}_{i}\right\rangle\left\langle\theta_{i}\right\rangle}\right]\frac{\nu}{L_{\Delta}^{2}\sqrt{\left\langle\left|S^{(1)}_{ij}\right|^{2}\right\rangle}}.
\end{equation}
As shown above in this sub-section, the term in square brackets in
equation (\ref{eq:alpha}) is constant in time. Figure 7 suggests that
$L_{\Delta}$ and $l_{\lambda}^{(1)}$ have the same dependence on time
but not the same dependence on viscosity.
Therefore, the time dependence of
$\left(L_{\Delta}^{2}\sqrt{\left\langle\left|S^{(1)}_{ij}\right|^{2}\right\rangle}\right)$
is the same as the time dependence of
$\left(\left(\eta^{(1)}\right)^{2}\left(\tau_{\eta}^{(1)}\right)^{-1}\text{Re}^{(1)}_{\lambda}\right)\sim
\text{Re}^{(1)}_{\lambda}$. For cases F1 and F2, the reference field,
and therefore $\text{Re}^{(1)}_{\lambda}$ are statistically steady,
and it therefore follows from equation (\ref{eq:alpha}) that $\alpha$
is a time-constant during the similarity regime. For the same reason,
$\alpha$ is a time-constant after $\tau=9.4$ in the similarity regime
of case F3 because this is when the reference flow reaches the
statistically steady state. During $\tau\in[7.5,9.4]$ for F3,
$1/\text{Re}^{(1)}_{\lambda}$ decreases monotonically from $0.0187$ to
$0.0156$ as shown in figure \ref{fig:reynoldsTaylorF3}. This $20\%$
decrease is small compared to the variations of
$1/\text{Re}^{(1)}_{\lambda}$ at normalised times $\tau$ smaller than
$7.5$ and results in a small decrease of $\alpha$ in the corresponding
time period (i.e. a slow increase of $1/\alpha$, as shown in figure
3). Therefore, $\alpha$ can be considered to be approximately constant
in the similarity period $\tau\in[7.5,12.5]$ of F3.

We have seen at the end of subsection \ref{sec:Mechanisms of the uncertainty evolution} and
  figure \ref{fig:P-Epsilon} that $\alpha = \left\langle
  P_{\Delta}\right\rangle/\left\langle\varepsilon_{\Delta}\right\rangle$
  seems to be independent of viscosity but we also noted two
  paragraphs above that $\beta$ is not. The dependencies on viscosity
  of $\beta\nu$ and
  $\left(L_{\Delta}^{2}\sqrt{\left\langle\left|S^{(1)}_{ij}\right|^{2}\right\rangle}\right)$
  in equation (\ref{eq:alpha}) must therefore be the same and cancel
  each other.


Substituting equations (\ref{eq:approximation ODE
    of production}) and equation (\ref{eq:alpha}) into equation
  (\ref{eq:first simplification}), we obtain
\begin{equation}
	\label{eq:final approximation}
	\frac{\mathrm{d}}{\mathrm{d} t}\left\langle E_{\Delta}\right\rangle=\Gamma\sqrt{\left\langle\left|S^{(1)}_{ij}\right|^{2}\right\rangle}\left\langle E_{\Delta}\right\rangle,
\end{equation}
where
\begin{equation}
	\label{eq:coefficient ODE of production}
	\Gamma=-2\frac{1-\alpha}{\beta}\sum_{i=1}^{3}\left\langle\gamma^{(1)}_{i}\right\rangle\left\langle\theta_{i}\right\rangle .
\end{equation}
This is a general rewriting of equation (\ref{eq:first
  simplification}) with particularly interesting consequences for the
similarity regime when $\alpha$, $\beta$,
$\left\langle\gamma^{(1)}_{i}\right\rangle$ and
$\left\langle\theta_{i}\right\rangle$ are constant in
time. The dimensionless coefficient $\Gamma$
  defined by equation (\ref{eq:coefficient ODE of production}) is
  therefore constant in time during the similarity regime but may
  depend on Reynolds number (i.e. viscosity) via the dependence of
  $\beta$ on Reynolds number.

Looking at equation (\ref{eq:final approximation}), an exponential
growth of $\left\langle E_{\Delta} \right\rangle$ with a well-defined
Lyapunov exponent $\lambda$ can be derived during the similarity
regime because $\Gamma$ is constant in time:
\begin{equation}
\label{eq:lambdacoefficient ODE of production}
2\lambda=\Gamma\sqrt{\left\langle\left|S^{(1)}_{ij}\right|^{2}\right\rangle}=\frac{1}{\sqrt{2}}\Gamma\tau_{\eta}^{-1}.
\end{equation}
The exponential growth of average uncertainty energy is, therefore, a
consequence of similarity. How similarity (time-independent $\alpha$,
$\beta$, $\left\langle \theta_{i} \right\rangle$ and self-similar
evolution of the uncertainty spectrum in terms of $\left\langle
E_{\Delta} \right\rangle$ and $L_{\Delta}$) may be a consequence of
the presence of a strange attractor is, however, beyond this paper's
scope but the question is now posed for future investigations.

The dimensionless coefficient $\Gamma$ obtained from equation
(\ref{eq:coefficient ODE of production}) and the Lyapunov exponent
directly obtained from equation (\ref{eq:exponential growth of
  uncertainty}) are plotted in figure \ref{fig:gamma}: for all cases
F1, F2 and F3, $\Gamma$ is about constant in the time range where
exponential growth is present. The actual value of $\Gamma$ in this
time range is the same for F1 and F2 but it is different for F3 which
has a lower Reynolds number. The scaling
$\lambda\tau_{\eta}\sim\Gamma(\text{Re})$ suggests that the Lyapunov
exponent may not scale with the Kolmogorov time $\tau_{\eta}$ (as
claimed by \citet{ruelle1979microscopic}) if $\Gamma$ depends on
Reynolds number, which it may do on account of a Reynolds number
dependence of $\beta$. The coefficient $\Gamma$,
  as well as the Lyapunov exponent, are also plotted in figure
  \ref{fig:Mohan} to compare with previous data by
  \citet{mohan2017scaling}. The ratio of $\Gamma$ values in the F1
and F3 cases is
$\Gamma_{\text{F1}}/\Gamma_{\text{F3}}=1.29$
  during the exponential growth time range, while
  $\beta_{\text{F3}}/\beta_{\text{F1}}=1.25$ in the same regime. The
data of \citet{mohan2017scaling} lead to
$\Gamma_{\text{F1}}/\Gamma_{\text{F3}} \approx
  1.30$ purely on the basis of the Reynolds numbers of F1 and F3 (see
figure \ref{fig:Mohan}). This confirms the hypothesis that the
different values of $\Gamma$ in F1 and F2 on the one hand and F3 on
the other are caused by the difference in Reynolds number and nothing
else.

The regime of approximate constancy of $\Gamma$ is followed by a time
range $\tau\in[2.9,6.5]$ where $\Gamma$ appears
to decay exponentially in the F1 and F2 cases (it is not clear whether
such a range does or does not exist in the F3 case), see figure
\ref{fig:gamma}. Specifically, the exponential curve fit gives
$\Gamma=0.73\exp(-1.26(\tau-2.9))$. Using
$\left\langle E_{\text{tot}}\right\rangle$ and $\left\langle
T^{(1)}\right\rangle_{t}$ to non-dimensionalise equation
(\ref{eq:final approximation}),
we write
\begin{equation}
\label{eq:non-dimensionalized approximation ODE of production}
\frac{\mathrm{d} }{\mathrm{d}\tau}\frac{\left\langle
	E_{\Delta}\right\rangle}{\left\langle
	E_{\text{tot}}\right\rangle}=\Gamma\left\langle
T^{(1)}\right\rangle_{t}\sqrt{\left\langle\left|S^{(1)}_{ij}\right|^{2}\right\rangle}\frac{\left\langle
	E_{\Delta}\right\rangle}{\left\langle 
	E_{\text{tot}}\right\rangle} . 
\end{equation}
For statistically stationary cases F1 and F2 we find
$\left\langle
  T^{(1)}\right\rangle_{t}\sqrt{\left\langle\left|S^{(1)}_{ij}\right|^{2}\right\rangle}=12.77\pm0.56$
in the time range $\tau\in[2.9,6.5]$. Therefore,
equation (\ref{eq:non-dimensionalized approximation ODE of
  production}) and our fitting of $\Gamma$ imply
$\left\langle E_{\Delta}\right\rangle/\left\langle
  E_{\text{tot}}\right\rangle\sim\exp(-9.32\exp(-1.26(\tau-2.9)))$,
which is approximately consistent with the direct curve fitting in the
inset of figure \ref{fig:time evolution of uncertinty
  compare}. Eventually $\Gamma$ tends to $0$ and the average
uncertainty energy stops growing in all cases F1, F2 and F3.

\subsubsection{Scaling of the Lyapunov exponent during similarity}

Our analysis in section \ref{sec:From similarity
    to exponential growth} and the data of \citep{mohan2017scaling}
  presented in figure \ref{fig:Mohan} question the view that $\lambda$
  scales with $\tau_{\eta}$ \citep{ruelle1979microscopic}.  If
$\lambda$ does not scale with $\tau_{\eta}$ which is the smallest
Lagrangian time scale of the turbulence, it may scale with
$\tau_{E}=\eta/U$, the shortest Eulerian time scale of the turbulence
\citep{tennekes1975eulerian}, in which case
$\lambda\tau_{\eta}\sim\tau_{\eta}/\tau_{E}\sim\text{Re}_{\lambda}^{1/2}$. The
data of \citet{mohan2017scaling} in figure \ref{fig:Mohan} suggests
that $\lambda$ grows faster than $\tau_{\eta}^{-1}$ but slower than
$\tau_{E}^{-1}$ as Reynolds number increases, perhaps $\lambda \sim
\tau_{\eta}^{-(1-c)/2}\tau_{E}^{-(1+c)/2}$, i.e. $\lambda
\tau_{\eta}\sim \text{Re}_{\lambda}^{(1+c)/4}$, where $c\in(-1,1]$. In
  fact, the results of \citet{mohan2017scaling} suggest that $\lambda
  \tau_{\eta} \sim\text{Re}_{\lambda}^{(1+c)/4}$ where the most likely
  values of $c$ are between $0$ and $1/3$. The large scale random
  sweeping of the smallest eddies represented in the Eulerian time
  scale $\tau_E$ appears to influence the growth of uncertainty even
  though the uncertainty exists only at the the smallest scales during
  the chaotic exponential growth. Interestingly, this large-scale
  random-sweeping effect is reflected in the decreasing dependence of
  $\beta$ on Reynolds number (see equations (\ref{eq:coefficient ODE
    of production}) and (\ref{eq:lambdacoefficient ODE of
    production})) which implies that $\left\langle
  P_{\Delta}\right\rangle$ should be increasingly dominated by
  $\left\langle P_{\Delta}\right\rangle_{\text{Fluc}}$ rather than
  $\left\langle P_{\Delta}\right\rangle_{\text{Ave}}$ in equation
  (\ref{eq:reynolds decomposition of production}) as Reynolds number
  increases. There seems to be a relation between
    large-scale random sweeping and uncertainty production, and in
    particular between random sweeping and the way that compression
    and stretching affect average uncertainty production either
    through average compression/stretching rates or through the
    correlations of their fluctuations with uncertainty energy
    fluctuations in specific stretching/compressive directions. A
    Lagrangian or some combined Eulerian-Lagrangian description of
    uncertainty (e.g. see \citet{boffetta1997predictability}) as
    advocated by \citet{leith1972predictability} in the introduction 
     might have advantages over
    the present purely Eulerian approach as it may naturally account
    for the large-scale sweeping's effect on uncertainty and thereby
    return a reduced average uncertainty production. The large-scale
    sweeping's effect on uncertainty might also have some relation
    with the error located in positions of local flow structures
    that \citet{boffetta1997predictability} identified.

	\begin{figure}
	\centering 
	\subfigure[case F1]{
		\label{fig:PDF case1}
		\includegraphics[width=0.49\textwidth]{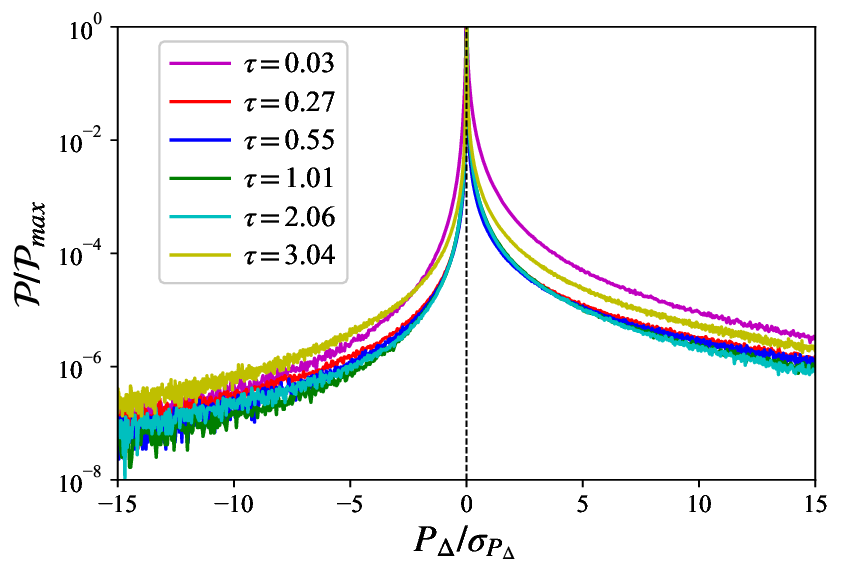}}
	\subfigure[case F2]{
		\label{fig:PDF case2}
		\includegraphics[width=0.49\textwidth]{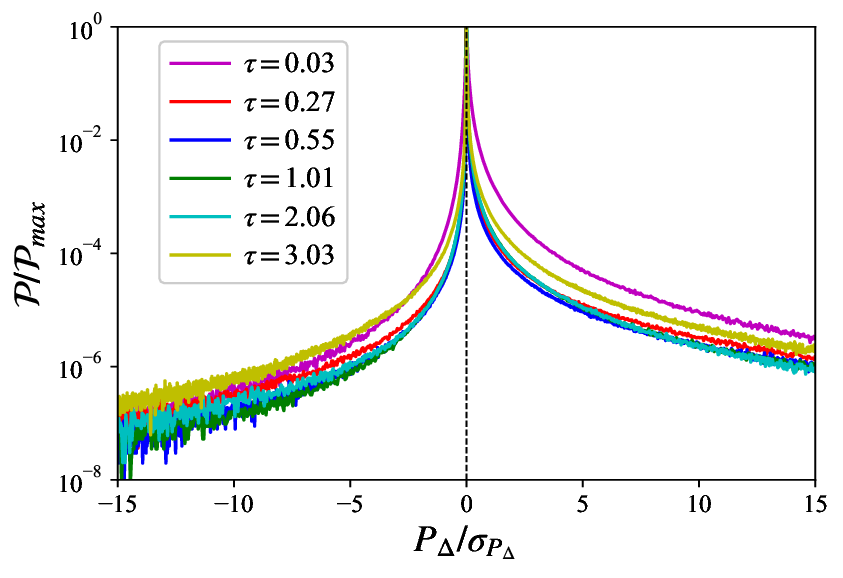}}
	\subfigure[case F3]{
		\label{fig:PDF case3}
		\includegraphics[width=0.49\textwidth]{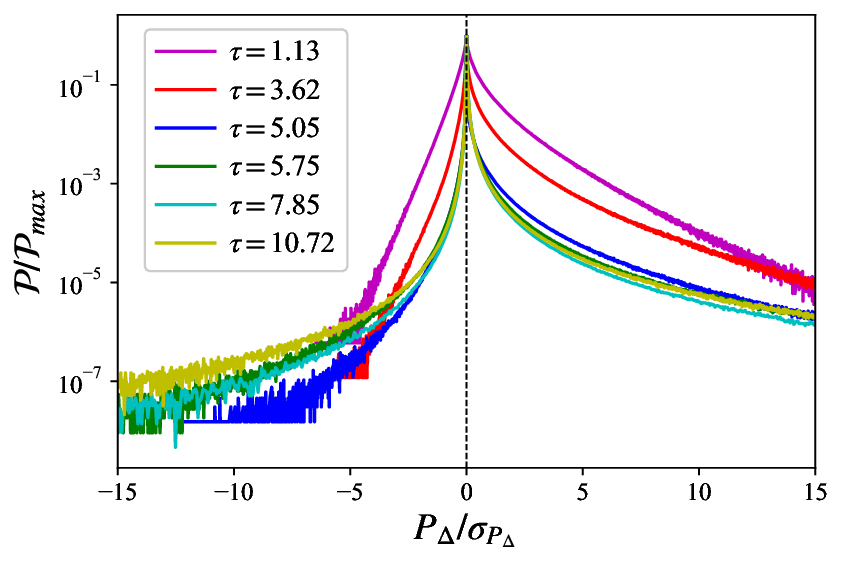}}
	\caption{Early-time evolution of PDFs of $P_{\Delta}$ for different cases. PDFs are plotted versus $P_{\Delta}/\sigma_{P_{\Delta}}$ where $\sigma_{P_{\Delta}}$ is the standard deviation of $P_{\Delta}$, defined as $\sigma^{2}_{P_{\Delta}}\equiv\int_{P_{\Delta\text{min}}}^{P_{\Delta\text{max}}}(P_{\Delta}-\left\langle
		P_{\Delta}\right\rangle)^{2}\mathcal{P}(P_{\Delta})\mathrm{d}P_{\Delta}$.} 
	\label{fig:PDF early} 
\end{figure}

\begin{figure}
	\centering 
	\subfigure[case F1]{
		\label{fig:PDF late case1}
		\includegraphics[width=0.49\textwidth]{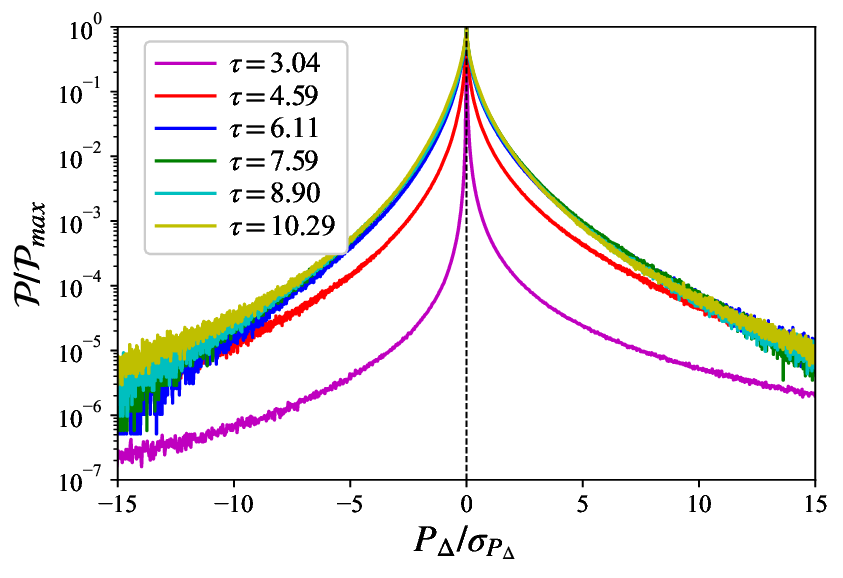}}
	\subfigure[case F2]{
		\label{fig:PDF late case2}
		\includegraphics[width=0.49\textwidth]{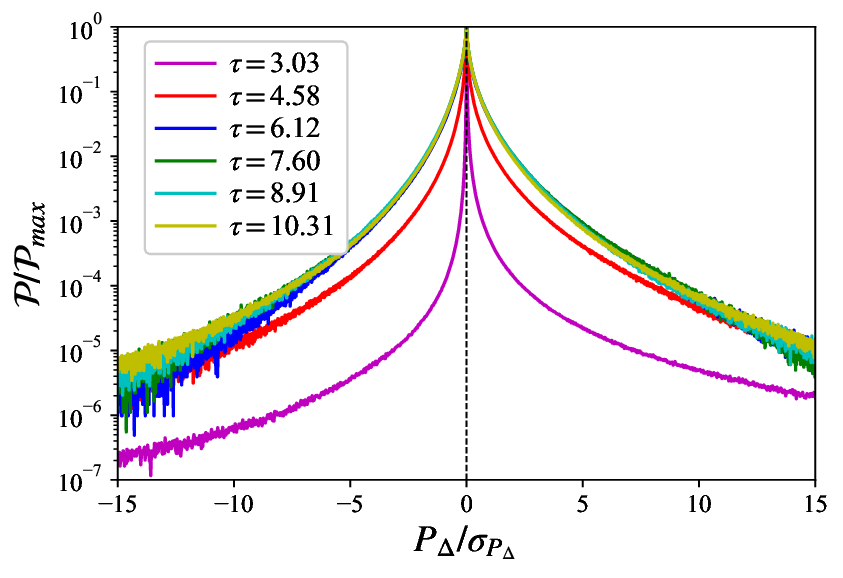}}
	\subfigure[case F3]{
		\label{fig:PDF late case3}
		\includegraphics[width=0.49\textwidth]{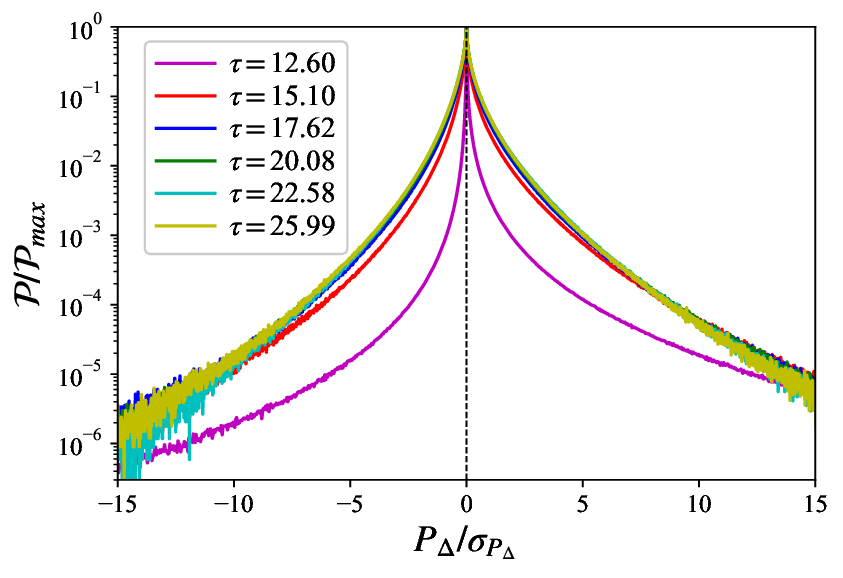}}
	\caption{Evolution of PDFs of $P_{\Delta}$ after the similarity
		regime for different cases. PDFs are plotted versus
		$P_{\Delta}/\sigma_{P_{\Delta}}$ where
		$\sigma_{P_{\Delta}}$ is the standard deviation of
		$P_{\Delta}$.}
	\label{fig:PDF late} 
\end{figure}

\begin{figure}
	\centering 
	\subfigure[case F1]{
		\label{fig:skewness case1}
		\includegraphics[width=0.49\textwidth]{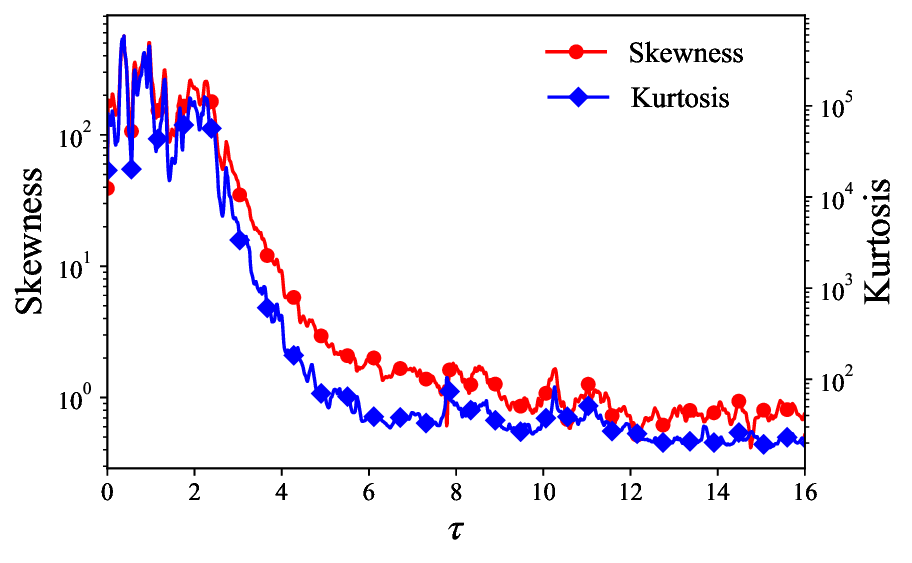}}
	\subfigure[case F2]{
		\label{fig:skewness case2}
		\includegraphics[width=0.49\textwidth]{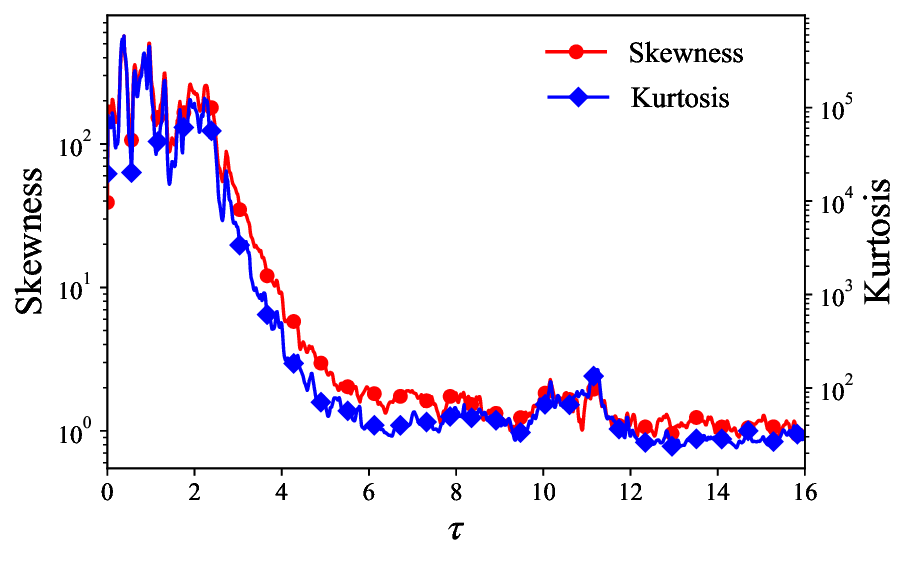}}
	\subfigure[case F3]{
		\label{fig:skewness case3}
		\includegraphics[width=0.49\textwidth]{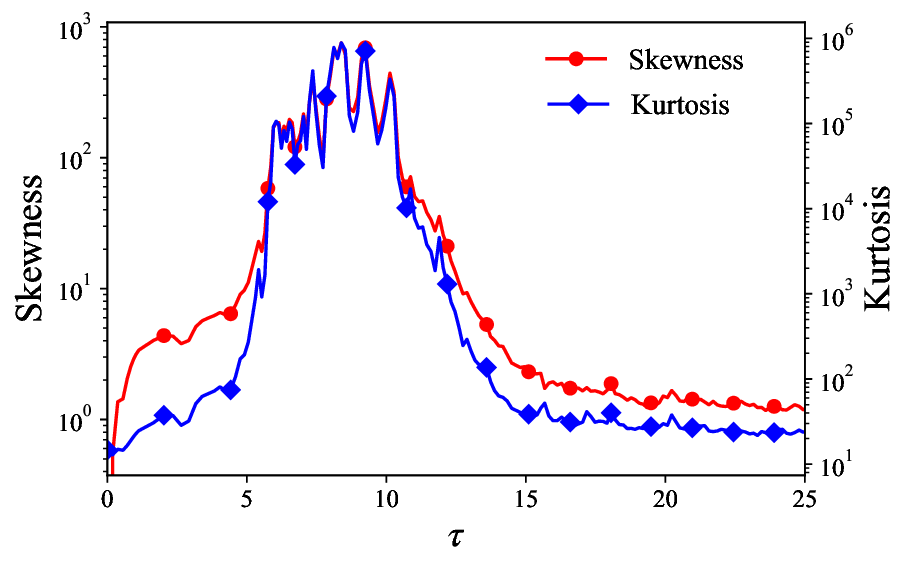}}
	\caption{Time evolution of the sample kurtosis and skewness of PDF for different cases, which is defined as $K=\frac{\int_{P_{\Delta\text{min}}}^{P_{\Delta\text{max}}}(P_{\Delta}-\left\langle P_{\Delta}\right\rangle)^{4}\mathcal{P}(P_{\Delta})\mathrm{d}P_{\Delta}}{\sigma_{P_{\Delta}}^{4}}$ and $S=\frac{\int_{P_{\Delta\text{min}}}^{P_{\Delta\text{max}}}(P_{\Delta}-\left\langle P_{\Delta}\right\rangle)^{3}\mathcal{P}(P_{\Delta})\mathrm{d}P_{\Delta}}{\sigma_{P_{\Delta}}^{3}}$.}
	\label{fig:skewness} 
\end{figure}

\subsection{\label{sec:The probability distribution of the uncertainty production}The probability distribution of the uncertainty production}

Even though the average uncertainty production rate is positive, the most likely value of $P_{\Delta}$ is zero at all times. In figures \ref{fig:PDF early} and \ref{fig:PDF late} we plot instantaneous probability density functions (PDF) of $P_{\Delta}$ sampled through all space and we examine how these PDFs evolve with time. An immediate observation is that the PDFs of $P_{\Delta}$ do not seem to match a well-known standard distribution (e.g. Gaussian, exponential, power-law) at any time and for any case F1, F2 and F3.  Another immediate observation is that the early time PDFs of $P_{\Delta}$ for F3 differ from those for F1 and F2 as their tails on the negative side are much shorter than on the positive side. These are times when the F3 reference flow is not statistically stationary.

Given that the most likely value of $P_{\Delta}$ is $P_{\Delta}=0$, the non-zero values of $\left\langle P_{\Delta}\right\rangle$ result from the positive skewnesses and the heavy tails of these PDFs (see figures \ref{fig:PDF early} and \ref{fig:PDF late}). The positive skewness and heavy tails, i.e. high kurtosis, set in from very early times and reveal an intermittent spatial distribution of co-existing uncertainty generation and depletion events where high generation events are more intense than high depletion events.

This spatial intermittency becomes increasingly acute and increasingly favourable to uncertainty generation rather than depletion events as the skewness and the kurtosis grow to extremely high positive values which fluctuate around a constant during the chaotic exponential growth in all F1, F2 and F3 cases (see figure \ref{fig:skewness}). This happens within the similarity regime where $\alpha$, $\beta$ and $\theta_i$ are constant and the uncertainty spectrum is self-similar if scaled with $\left\langle E_{\Delta}\right\rangle$ and $L_{\Delta}$. In fact, as shown in figure \ref{fig:PDF early}, the PDFs of $P_{\Delta}$ also approximately collapse during the time range of extreme skewness and kurtosis if normalised by the PDF's maximum value and standard deviation. During this time range where similarity and exponential uncertainty growth coexist, the kurtosis and the skewness fluctuate around $10^{5}$ and $200$ respectively, suggesting that $\left\langle P_{\Delta} \right\rangle$ is predominantly determined by rare yet powerful events of uncertainty generation and depletion.

After the similarity and chaotic growth stage, both the skewness and the kurtosis of the PDFs continuously decrease with time indicating that more points in the flow participate in the uncertainty generation and depletion and in the overall value of $\left\langle P_{\Delta}\right\rangle$. The way these PDFs lead to the average values of $P_{\Delta}$ is subtle. The long time saturation value of $\left\langle P_{\Delta}\right\rangle$ is zero for F1 and non-zero for F2, yet the long time PDFs of $P_{\Delta}$ are similar in both cases, as are the long time values of kurtosis and skewness.

\section{\label{sec:Conclusion}Conclusion}

In the present work, we obtained the evolution equation (\ref{eq:total uncertainty equation}) for the average uncertainty energy $\left\langle E_{\Delta} \right\rangle (t)$ in three-dimensional, incompressible and periodic/homogeneous Navier-Stokes turbulence. The average uncertainty energy evolves because of internal production, dissipation and external input/output of uncertainty. The internal production of uncertainty is a transfer from the correlation between the reference and perturbed fields to the average uncertainty energy and is determined by the eigenvalues of reference field's strain rate tensor and the distribution of uncertainty energy along its three eigenvectors. As shown by equation (\ref{eq:Production in principal axe}), stretching events decrease uncertainty while the compression events increase uncertainty.

We used DNS of periodic Navier-Stokes turbulence to study the gradual
decorrelation process of two initially highly correlated flows. Three
different DNS were run, F1, F2 and F3: two where the perturbation is
seeded to a statistically stationary turbulence and where the forcing
does (F1) or does not (F2) contribute directly to the progressive
decorrelation between the reference and perturbed fields; and one (F3)
where the reference and perturbed fields are both initially very weak
and grow together to eventually become statistically stationary
without the external forcing contributing directly to their gradual
decorrelation. In all three cases and at times when $\left\langle
E_{\Delta} \right\rangle (t)$ is still small, a similarity time-range
was found where the growth of the uncertainty spectrum is self-similar
if scaled by $\left\langle E_{\Delta} \right\rangle (t)$ and the
characteristic length $L_{\Delta} (t)$ of uncertainty, and where all
the following quantities are constant in time: (i) the ratio $\alpha$
of average uncertainty dissipation to average uncertainty production,
(ii) the ratio $\beta$ characterising how much of the average
uncertainty production rate is accountable to the average around which
it fluctuates in space, and (iii) the distribution of uncertainty
energy in the three eigen-directions of the reference field's strain
rate tensor. These three similarity constancies and the constancy in
time of the three average eigenvalues of the reference field's strain
rate tensor imply an exponential growth in time for $\left\langle
E_{\Delta}\right\rangle$ with Lyapunov exponent $\lambda \sim \Gamma
\tau_{\eta}^{-1}$. The dimensionless coefficient $\Gamma$ is given by
equation (\ref{eq:coefficient ODE of production}) and grows with
Reynolds number because $\beta$ decreases with Reynolds number. This
exponential growth for $\left\langle E_{\Delta}\right\rangle$ is
observed in the earlier part of the time range of the similarity
regime when the PDF of $P_{\Delta}$ collapses for different times if
scaled by its maximum value and standard deviation. As a result, the
kurtosis and skewness of this PDF are about constant in this time
range. In fact, the value of this constant kurtosis is extremely large
indicating extreme intermittency of $P_{\Delta}$. The value of the
constant skewness is also large and positive indicating that rare high
uncertainty generation events are more intense than rare high
uncertainty depletion events. The average value of $P_{\Delta}$ is
controlled by this intermittency in this time range. Note that the
most probable value of $P_{\Delta}$ is zero at all times.

During the chaotic exponential growth regime, $L_{\Delta}$ versus the
Taylor length $l_{\lambda}$ of the reference flow is about the
constant. In agreement with previous observations
\citep{mohan2017scaling}, the Lyapunov exponent does not scale with
the Kolmogorov time $\tau_{\eta}$, but it also does not scale with the
smallest Eulerian time scale $\tau_{E}$
\citep{tennekes1975eulerian}. It appears to depend on both as $\lambda
\sim \tau_{\eta}^{-(1-c)/2} \tau_{E}^{-(1+c)/2}$ with $c$ between $0$
and $1/3$, implying that large scale random sweeping of the smallest
length-scales influences the growth of uncertainty even though
uncertainty only exists in the smallest eddies in the time range of
chaotic exponential growth.

The chaotic growth time-range is followed by a time-range in the F1
and F2 cases where $\Gamma$ decays exponentially and $\left\langle
E_{\Delta}\right\rangle$ grows as an exponential of an exponential. In
turn, this exponential of exponential time-range may be followed by a
linear time range in the F1 case consistently with previous DNS
studies \citep{berera2018chaotic,boffetta2017chaos}, but not in the F2
case, at least for our present DNS Reynolds numbers. The linear growth
of uncertainty seems to be sensitive to the direct presence (F1) or
absence (F2) of external forcing in the evolution of $\left\langle
E_{\Delta}\right\rangle$. We did not detect a linear time growth of
$\left\langle E_{\Delta}\right\rangle$ in F3 either, however the F3
Reynolds number is even lower.

Finally, the exponential growth of $\left\langle
E_{\Delta}\right\rangle$ is usually attributed to the presence of a
strange attractor whereas it has been obtained here from
similarity. Future research should attempt to shed light on the
relations between similarity and strange attractors, and on how
similarity may be a consequence of the presence of a such an attractor
and underlying chaos. Future research may also
  consider how this paper's approach to uncertainty in homogeneous
  turbulence can be extended to a wider range of turbulent flows. In
  general, the governing equation for Navier-Stokes uncertainty is
  (\ref{eq:singlepoint uncertainty equation}) rather than
  (\ref{eq:total uncertainty equation}). Hence, turbulent as well as
  viscous diffusion and also pressure effects will need to be taken
  into account explicitely in the evolution of uncertainty. Various
  boundary conditions and errors on boundary conditions in case of
  complex turbulent flows will also be an issue, not to mention
  various body forces and the presence in many turbulent flows of
  turbulent/non-turbulent or turbulent/turbulent or other
  (e.g. density) interfaces. The identification of local compression
  and stretching events as key to the development of uncertainty means
  that future prediction methods may benefit from ways to detect early
  such events so as to concentrate maximum accuracy on the compression
  ones and less accuracy on the stretching ones. However, the roles of
  all the other aforementioned effects should not be understimated and
  future research is needed to show whether they are subdominant or
  not and in which flows.



\backsection[Acknowledgements]{Jin Ge acknowledges financial support
  from the China Scholarship Council. We are grateful for the access
  to the computing resources supported by the Zeus supercomputers
  (Mésocentre de Calcul Scientifique Intensif de l'Université de
  Lille).}

\backsection[Funding]{This research received no specific grant from any funding agency, commercial or not-for-profit sectors. }

\backsection[Declaration of interests]{The authors report no conflict of interest.}



\appendix

\section{Sensitivity of the uncertainty energy to the initial perturbation}\label{app:Sensitivity of the uncertainty energy to the initial perturbation}
To investigate the sensitivity of the evolution of average uncertainty
energy to the initial perturbation, a series of simulations have been
executed, of which the configurations are presented in table
\ref{tab:num config}. By checking the evolution of the average
uncertainty energy, the influence of the perturbed range (cases
``standard", ``K07K08" and ``K08K09") and of the amplitude (cases
``standard" and ``Amp01") of the initial perturbation is
investigated. During the similarity period, the changes in the
amplitude and the perturbed range have very little effect on the
evolution of the average uncertainty energy, other than giving the
evolution an offset (explained below). At late times, the difference
between average uncertainty energies induced by different initial
perturbations becomes more obvious for F1 where the external forcing
causes an eventual decorrelation between the perturbed and the
unperturbed velocity fields.
\begin{table}
	\centering
	\begin{tabular}{ccc}
		Case&$\left\langle E_{\Delta}(t_{0})\right\rangle/\left\langle E_{\text{tot}}(t_{0})\right\rangle$&Perturbed range\\
		\midrule
		Standard (F1 or F2)&$8.077\times10^{-6}$&$\left[0.9k_{max},1.0k_{max}\right]$\\
		K08K09&$8.077\times10^{-6}$&$\left[0.8k_{max},0.9k_{max}\right]$\\
		K07K08&$8.077\times10^{-6}$&$\left[0.7k_{max},0.8k_{max}\right]$\\
		Amp01&$8.077\times10^{-7}$&$\left[0.9k_{max},1.0k_{max}\right]$\\
	\end{tabular}
	\caption{Numerical configurations for different cases. The two
		standard cases correspond to F1 and F2 in the
		manuscript. There are two cases K08K09, one for F1 and one
		for F2, and similarly for cases K07K08 and Amp01. For the
		standard F1 and F2 cases, the initial perturbations are
		generated randomly under constraints (1), (2) and (3) mentioned in section \ref{sec:Numerical steups}, but
		for the other six cases the initial perturbations are
		generated partially randomly under constraints (1) and (2) in order to precisely control the initial uncertainty energy.}
	\label{tab:num config}
\end{table}

\begin{figure}
	\centering  
	\subfigure[case F1]{
		\label{Fig.sub.1}
		\includegraphics[width=0.49\textwidth]{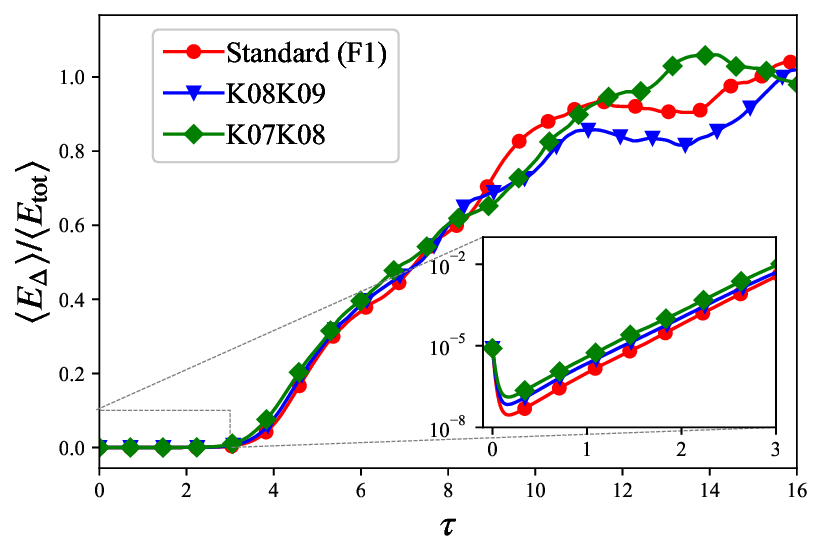}}
	\subfigure[case F2]{
		\label{Fig.sub.2}
		\includegraphics[width=0.49\textwidth]{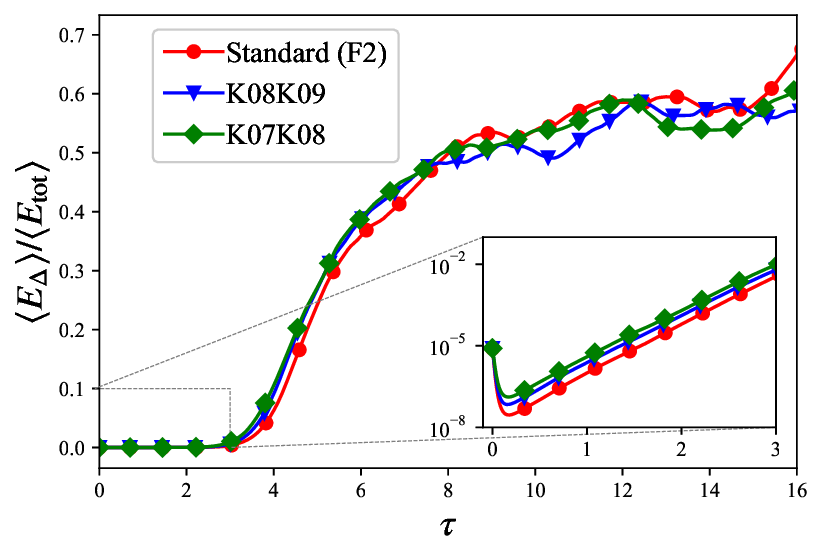}}
	\caption{Time evolution of average uncertainty energy with
		different perturbed wavenumber range. Inset: the initial
		time evolution of average uncertainty energy in
		semilogarithmic plot.}
	\label{fig:growthcertaintyenergy_Perturb_range}
\end{figure}

Figure \ref{fig:growthcertaintyenergy_Perturb_range} presents the time
evolutions of the average uncertainty energy for different perturbed
wavenumber ranges. A higher wavenumber perturbed range implies higher
uncertainty dissipation rate for the seeded uncertainty at the
earliest times, which causes lower value of $\left\langle
E_{\Delta}\right\rangle/\left\langle E_{\text{tot}}\right\rangle$ at
very early times and during the similarity period. The effect appears
in the log-linear inset of figure
\ref{fig:growthcertaintyenergy_Perturb_range} as a vertical offset of
the curves for the different cases. The average uncertainty energy
grows exponentially in all three cases with the same Lyapunov
exponent. These different vertical offsets lead to slightly different
exit times from the similarity regime. The regime of exponential
growth is followed by what appears to be an exponential of exponential
regime, where the difference of wavenumber perturbed range has little
influence on the evolution of average uncertainty energy since the
lines in figure \ref{fig:growthcertaintyenergy_Perturb_range} are very
close to each other albeit with a persisting small offset.

\begin{figure}
	\centering  
	\subfigure[case F1]{
		\includegraphics[width=0.49\textwidth]{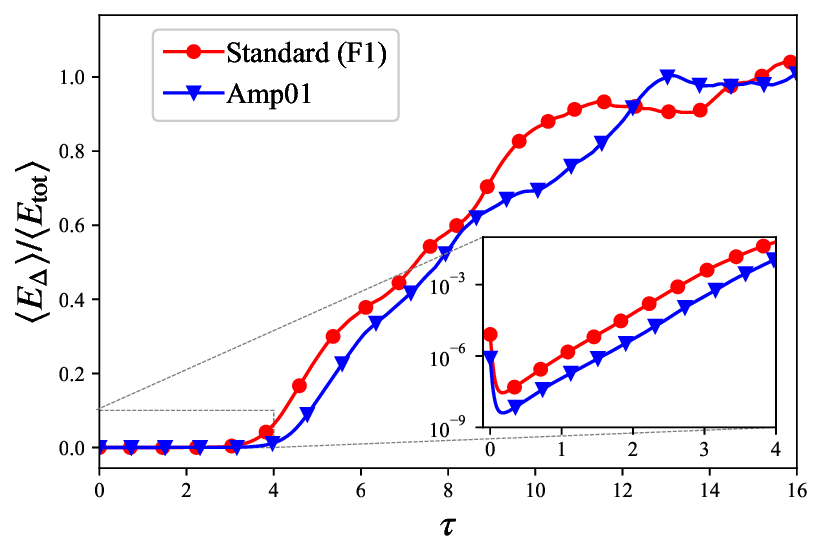}}
	\subfigure[case F2]{
		\includegraphics[width=0.49\textwidth]{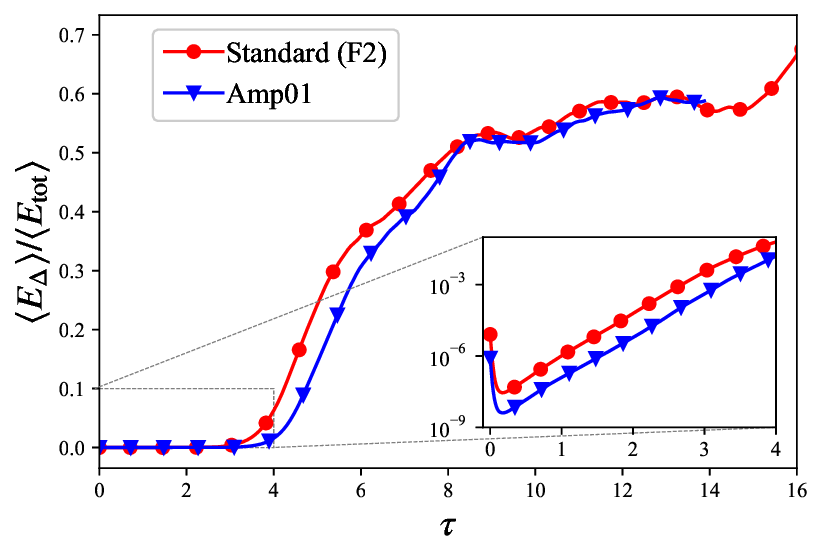}}
	\caption{Time evolution of average uncertainty energy with different initial uncertainty energy. Inset: the initial time evolution of average uncertainty energy in semilogarithmic plot.}
	\label{fig:growthcertaintyenergy_Amplitude}
\end{figure}

Figure \ref{fig:growthcertaintyenergy_Amplitude} presents the time
evolution of the average uncertainty energy for the different initial
uncertainty energy levels. As can be seen in the figure, the change in
the amplitude of initial perturbation has the same effect as the
change in the perturbed wavenumber range, i.e., no significant
influence on the evolution of uncertainty energy other than creating
an offset.

\begin{figure}
	\centering  
	\subfigure[case F1]{
		\includegraphics[width=0.49\textwidth]{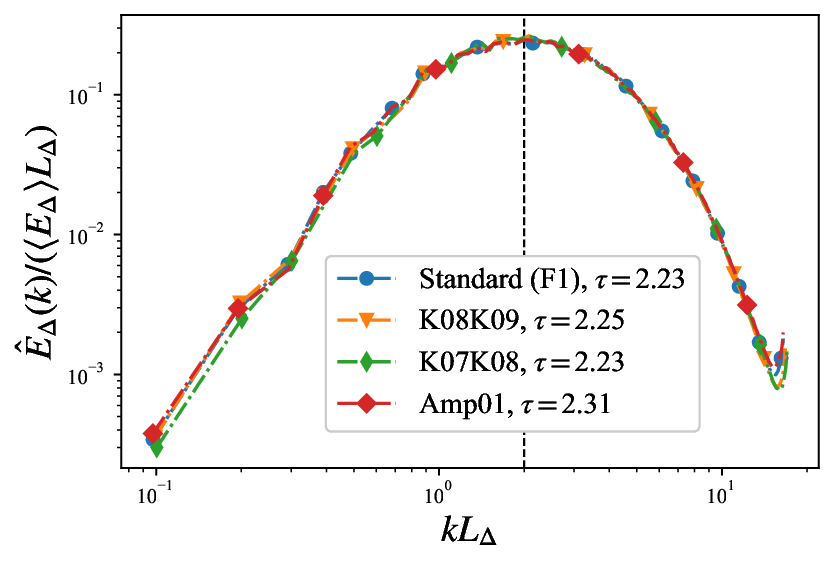}}
	\subfigure[case F2]{
		\includegraphics[width=0.49\textwidth]{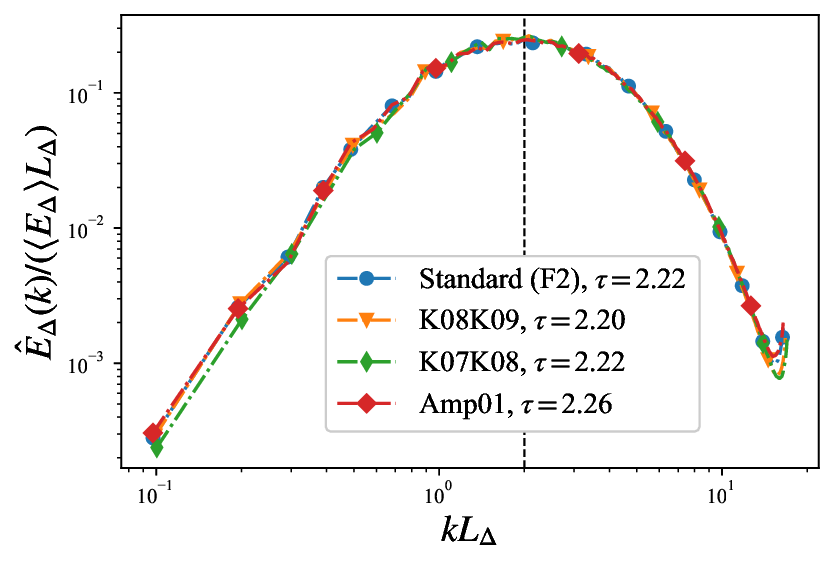}}
	\caption{Uncertainty energy spectra in the similarity regime. The spectra are normalized by $\left\langle E_{\Delta}\right\rangle$ and $L_{\Delta}$.}
	\label{fig:errorspectra_Perturb_range}
\end{figure}

We also checked the uncertainty spectra in the self-similar regime for
our various cases with different initial perturbations, as shown in
figure \ref{fig:errorspectra_Perturb_range}. All the self-similar
spectra with different initial perturbations collapse together.

As an overall conclusion, the early- and mid-time evolutions of the
average uncertainty energy are not very sensitive to the form and
amplitude of the initial perturbations, other than giving the
evolution an offset.

\section{\label{app:Reynolds-number dependence of the time range of the exponential
		regime}Reynolds-number dependence of the time range of the exponential
	regime}
\begin{figure}
	\centering  
	\subfigure[case F2 - case F4]{
		\label{fig:errorspectra_Perturb_range F2F4 a}
		\includegraphics[width=0.49\textwidth]{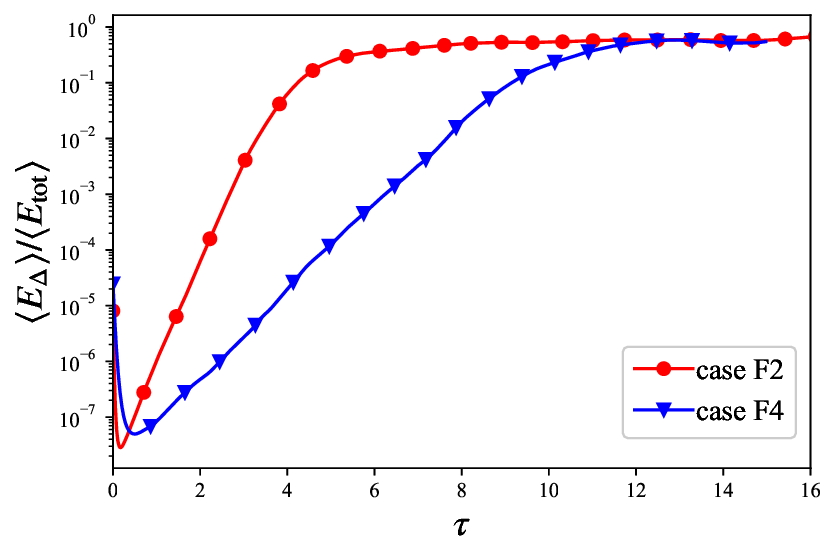}}
	\subfigure[case F3 - case F4]{
		\label{fig:errorspectra_Perturb_range F2F4 b}
		\includegraphics[width=0.49\textwidth]{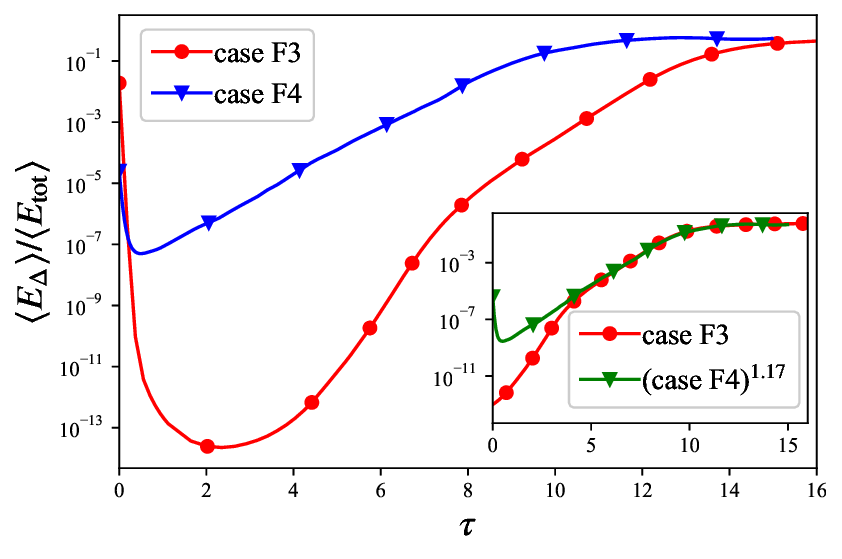}}
	\caption{Time evolution of average uncertainty energy in
		semilogarithmic plot. In the inset of (b), we plot
		$(\left\langle E_{\Delta}\right\rangle/\left\langle
		E_{\text{tot}}\right\rangle)^{1.17}$ for F4 and
		$(\left\langle E_{\Delta}\right\rangle/\left\langle
		E_{\text{tot}}\right\rangle)$ for F3 translated in the
		horizontal axis by 2.7 $\tau$-units to the left.}
	\label{fig:errorspectra_Perturb_range F2F4}
\end{figure}

\begin{table}
	\centering
	\begin{tabular}{lccccccccc}
		Case&$N^{3}$&$\nu$&$\left\langle\left\langle\varepsilon\right\rangle\right\rangle_{t}$&$\left\langle U\right\rangle_{t}$&$\left\langle L\right\rangle_{t}$&$\left\langle T_{0}\right\rangle_{t}$&$\left\langle\text{Re}\right\rangle_{t}$&$\left\langle\text{Re}_{\lambda}\right\rangle_{t}$&$\left\langle k_{\max}\eta\right\rangle_{t}$\\ [3pt]
		\midrule
		F4&$128^{3}$&0.0060&0.0996&0.598&1.197&2.003&119.2&56.7&1.61\\
	\end{tabular}
	\caption{Parameters of the reference flows for case F4.}
	\label{tab:main parameters2}
\end{table}
To investigate the relation between the time range of the exponential
regime and the Reynolds number, we have run another simulation which
has the same external forcing as F2 with initial perturbations which,
like standard F1, F2 and F3, obey the three constraints mentioned in section \ref{sec:Numerical steups}. Table \ref{tab:main parameters2} presents the
main parameters of this extra case F4, as well as cases F2/F3
discussed in the manuscript. As shown in table \ref{tab:main parameters} and table \ref{tab:main parameters2}, the Taylor Reynolds number of case F4 is close to that
of case F3. Figure \ref{fig:errorspectra_Perturb_range F2F4} presents
the growths of average uncertainty in a semilogarithmic plot. In
figure \ref{fig:errorspectra_Perturb_range F2F4 a} we compare the
evolution in cases F2 and F4. As can be seen in the figure, the
exponential regime in F4 is longer than in F2, and also has a slower
growth rate than F2, which is (see equation (\ref{eq:non-dimensionalized approximation ODE of production}))
\begin{equation}
	\label{eq:growth rate}
	\Gamma\left\langle
	T^{(1)}\right\rangle_{t}\sqrt{\left\langle\left|S^{(1)}_{ij}\right|^{2}\right\rangle}\sim\Gamma(\text{Re}_{\lambda})\cdot\text{Re}_{\lambda}.
\end{equation}
The lower Reynolds number case has a lower growth rate. Furthermore,
as shown in figure \ref{fig:uncertainty spectra}, the exit time from the
similarity regime corresponds to the moment when the velocities at the
largest wavenumbers become completely decorrelated,
i.e. $\hat{E}_{\Delta}(k_{max})=\hat{E}_{\text{tot}}(k_{max})$.
Therefore, as the Reynolds number increases, the energy spectrum's
inertial range also increases towards smaller scales, causing a
decreasing threshold value $\left\langle
E_{\Delta}\right\rangle/\left\langle E_{\text{tot}}\right\rangle$ that
needs to be overcome for the exit time from the exponential growth
regime.
As a result, the lower Reynolds number case has a longer time-range of
exponential growth.

In figure \ref{fig:errorspectra_Perturb_range F2F4 b} we compare the
exponential growths in cases F3 and F4. It is observed that cases F3
and F4 have similar exponential growth rates. The slight difference in
exponential growth rates is caused by the small difference in Reynolds
numbers. To verify this point, equation (\ref{eq:growth rate}) is
applied, along with the observation of \citet{mohan2017scaling} that $\Gamma(\text{Re}_{\lambda})\sim
\text{Re}_{\lambda}^{1/3}$. Therefore, we predict that the ratio of
exponential growth rates of F3 and F4 is $(63.8/56.7)^{4/3}=1.17$,
which is verified by our simulations as shown in the inset of figure
\ref{fig:errorspectra_Perturb_range F2F4 b}. Although cases F3 and F4
have the similar exponential growth rates, case F4 has a longer
exponential regime. This may have something to do with the fact that
F3 is not statistically stationary until $\tau = 9.3$ whereas F4 is
statistically stationary from the start of the perturbation.

\bibliographystyle{jfm}
\bibliography{jfm}

\end{document}